%% file: sample-sigconf.tex
\documentclass[manuscript]{acmart}
\AtBeginDocument{%
  }
 
\setcopyright{acmlicensed}
\copyrightyear{2018}
\acmYear{2018}
\acmDOI{XXXXXXX.XXXXXXX}
\acmConference[Conference acronym 'XX]{Make sure to enter the correct
  conference title from your rights confirmation email}{June 03--05,
  2018}{Woodstock, NY}
\acmISBN{978-1-4503-XXXX-X/2018/06}

\usepackage{tcolorbox}
\usepackage{booktabs}
\usepackage{tabularx}
\usepackage{array}

\begin{document}

\title[Supporting Habit Formation in Pre-Meal Photo Capture]{Supporting Habit Formation in Pre-Meal Photo Capture: A Longitudinal Study of Behavior Change Techniques}


\author{Jun Fang}
\affiliation{%
  \institution{Department of Computer Science and Technology, Tsinghua University}
  \country{China}
}
\email{fangy23@mails.tsinghua.edu.cn}
\orcid{0009-0001-2614-8674}

\author{Yueying Suo}
\affiliation{
  \institution{Sino-French Institute, Renmin University of China}
  \city{Beijing}
  \country{China}
}
\email{yueying_suo@ruc.edu.cn}
\orcid{0009-0007-5613-1316}

\author{Yuntao Wang}
\authornote{Corresponding author.}
\affiliation{%
  \institution{Key Laboratory of Pervasive Computing, Ministry of Education, Department of Computer Science and Technology, Tsinghua University}
  \country{China}
}
\email{yuntaowang@tsinghua.edu.cn}
\orcid{0000-0002-4249-8893}

\author{Susu Wang}
\affiliation{
  \institution{Beijing University of Chemical Technology}
  \city{Beijing}
  \country{China}
}
\email{19156131236@163.com}
\orcid{0009-0004-6002-7085}

\author{Xiaoyu YANG}
\affiliation{
  \institution{The Hong Kong University of Sciecne and Technology (Guangzhou)}
  \city{Guangzhou}
  \country{China}
}
\email{xyang058@connect.hkust-gz.edu.cn}
\orcid{0009-0000-6169-9669}

\author{Qiqi Liu}
\affiliation{%
  \institution{Department of Automation, Tsinghua University}
  \city{Beijing}
  \country{China}
}
\email{lqq24@mails.tsinghua.edu.cn}
\orcid{0009-0005-4250-9192}

\author{Chuangyu Ding}
\affiliation{
  \institution{Institut Polytechnique de Paris}
  \city{Palaiseau}
  \country{France}
}
\email{chuangyu.ding@polytchnique.edu}
\orcid{0009-0000-5205-8536}

\author{Yuanchun Shi}
\orcid{0000-0003-2273-6927}
\affiliation{%
  \institution{Key Laboratory of Pervasive Computing, Ministry of Education, Department of Computer Science and Technology, Tsinghua University}
  \country{China}
}
\affiliation{%
  \institution{Intelligent Computing and Application Laboratory of Qinghai Province, Qinghai University}
  \country{China}
}
\email{shiyc@tsinghua.edu.cn}

\renewcommand{\shortauthors}{Fang et al.}

\begin{abstract}

\begin{figure}[h]
    \centering
    \includegraphics[width=\linewidth]{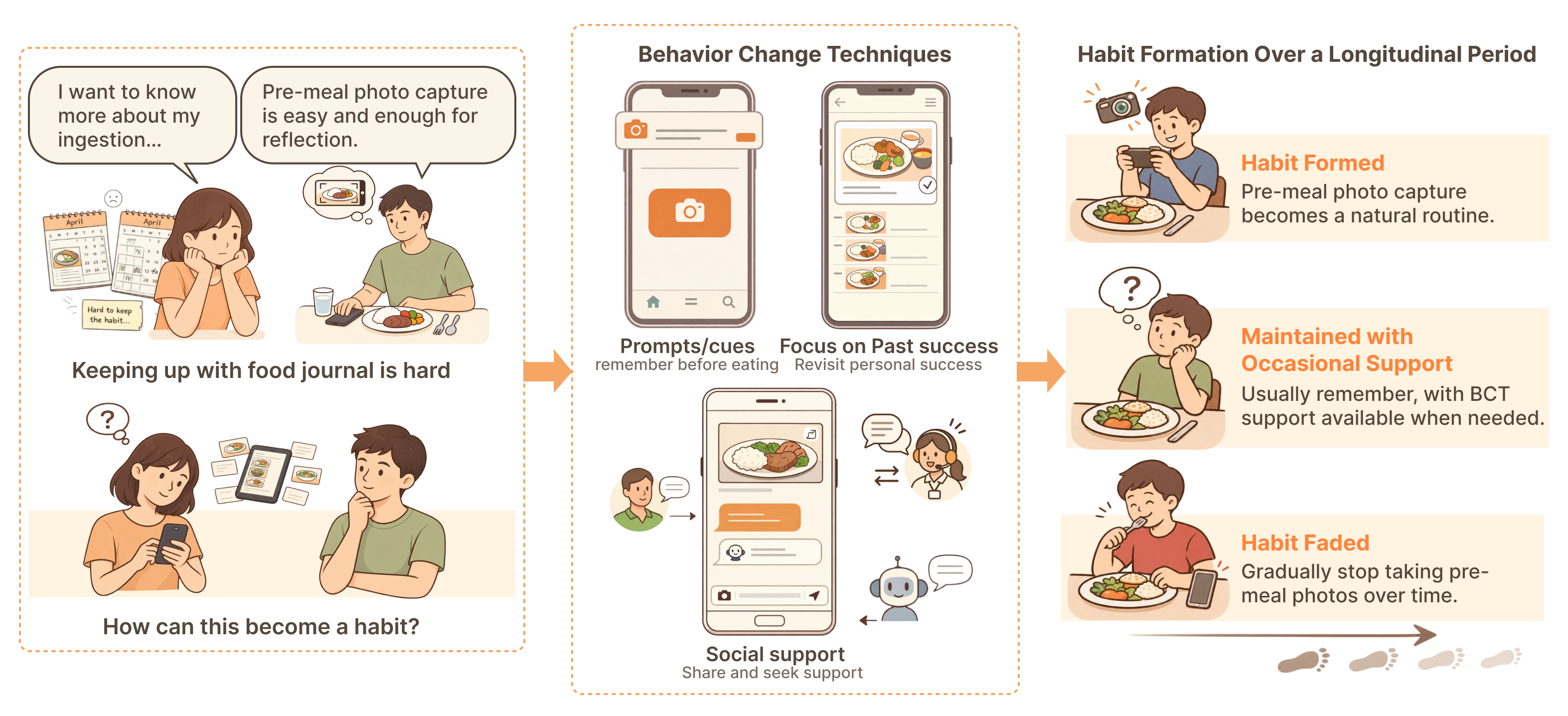}
    \caption{Overview of the habit formation in pre-meal photo capture. We examine how BCT-based supports selected for longitudinal evaluation, namely Prompts/cues, Focus on past success, and Social support, can help motivated users integrate this behavior into everyday routines. Their roles and users' support needs may follow different trajectories as capture practices and everyday contexts change over time.}
    \label{fig:teaser}
\end{figure}

Pre-meal photo capture is lightweight only with a smartphone, yet habit formation requires repeatedly initiating and completing the action within the narrow window before the first bite. We examine how supports based on behavior change techniques can shape capture adherence, perceived habit, and support needs over time. Using a participant-informed selection process, we implemented three smartphone supports targeting Prompts/cues, Focus on past success, and Social support. We evaluated them in a five-week field study with 54 participants, spanning support, withdrawal, and reinstatement, followed by interviews and a three-week follow-up. Across 4,945 main meals, 63.9\% were confirmed pre-meal captures, and adherence was broadly maintained across phases. Perceived automaticity and spontaneous remembering improved mainly during initial support. As capture became more familiar, participants combined spontaneous remembering with deliberate strategies, reduced active feature use while continuing to value support availability, and carried the practice forward through accumulated records, familiar tools, and chosen relationships. These findings frame BCT support for pre-meal photo capture as a configurable arrangement. Systems should preserve a predictable capture pathway while adapting when, how, and through whom support remains available.

\end{abstract}




\maketitle

\input{section/1_introduction}
\input{section/2_related_works_background}
\input{section/3_formative_selection}

\input{section/4_system_design}
\input{section/5_longitudinal_study}
\input{section/6_results}
\input{section/7_discussion}

\input{section/8_conclusion}

\bibliographystyle{ACM-Reference-Format}
\bibliography{ref}

\newpage
\appendix
\input{section/appendix.tex}

\end{document}

%% file: section/1_introduction.tex
\section{Introduction}




Food journals help people review their meals, identify dietary patterns, and reflect on and manage their eating practices in relation to personal goals~\cite{cordeiro2015rethinking,chung2019identifying,luo2019co}.
For people who wish to document their eating, integrating this practice into everyday routines can be valuable.
However, conventional manual logging requires users to enter and organize food information, which creates a recording burden~\cite{cordeiro2015barriers,koenig2022characteristics}.
Records focused on downstream goals may also omit important visual and contextual details~\cite{chung2019identifying,chung2024opportunities}.
Wearable sensing offers automated alternatives, but it can introduce physical and interaction burdens and raise concerns about comfort, privacy, and compatibility with daily routines~\cite{bedri2022fitnibble,lu2022understanding}.


Pre-meal photo capture offers a lightweight way to preserve food appearance and contextual cues~\cite{luo2021foodscrap,chung2024opportunities}.
Requiring only a smartphone, pre-meal photo capture enables accessible and portable recording in everyday settings, consistent with the goals of ubiquitous computing~\cite{koenig2022characteristics,weiser1991computer}.
Taking photos can encourage awareness of food at the time of capture~\cite{zepeda2008think}, while reviewing them can support reflection on recent eating patterns in relation to personal understandings of healthy eating~\cite{cordeiro2015rethinking,chung2019identifying}.

Despite the promise of pre-meal photo capture, establishing this practice as an everyday habit remains challenging.
Although taking a photo requires little interaction, forming a habit of timely and consistent capture requires repeatedly initiating and completing the action before eating and integrating it into daily meal routines~\cite{cordeiro2015rethinking,lee2024tailored,lally2010habits}.
Behavior change techniques (BCTs), defined as identifiable and irreducible intervention components, provide a broad, structured design space for developing such support~\cite{michie2013behavior,hekler2013mind}.
However, recent research has focused on downstream uses of food records and continued engagement with journal systems, providing limited insight into how the capture behavior itself becomes habitual~\cite{cordeiro2015rethinking,chung2019identifying,alberts2026designing}.
This gap motivates treating pre-meal photo capture itself as a habit-formation goal and establishing a specific basis for BCT selection.
Identifying promising BCTs for this goal requires considering evidence from HCI and behavioral science, constraints on smartphone implementation, and users' assessments of the likely usefulness of these techniques in everyday meal contexts~\cite{consolvo2009theory}.

Prior studies have evaluated prompting strategies for dietary recording~\cite{lee2024tailored,chen2024investigating} and compared the effects of cues and positive reinforcement on recording adherence and automaticity~\cite{stawarz2015beyond}.
However, evaluations of these preselected techniques offer limited insight into how habits form and how effects of BCT-based support change over time in everyday life.
Understanding these processes requires a longitudinal investigation of behavioral trajectories, experiences of habit formation, and support use across everyday meal contexts~\cite{rapp2023exploring,klasnja2011evaluate}.
These gaps motivate the following research questions:
\begin{itemize}
    \item \textbf{RQ1}: \textit{How do BCT-based supports shape adherence to pre-meal photo capture over time?}
    \item \textbf{RQ2}: \textit{How do the effects of these supports unfold over time, including during support withdrawal and reinstatement?}
    \item \textbf{RQ3}: \textit{What factors influence the impact of BCT-supported photo capture across everyday meal contexts, and how can these insights inform the design of future BCT-based systems?}
\end{itemize}


To address these gaps, we first conducted a formative study with 10 participants to select and refine BCTs that could be implemented on smartphones to support pre-meal photo capture.
This process yielded Prompts/cues, Social support, and Focus on past success, which we implemented in \textit{MealCapture}.
We then conducted a five-week, three-arm field study.
54 participants completed two weeks of support and one week of withdrawal, and 50 continued through two weeks of reinstatement.
We also conducted post-study interviews and administered a follow-up questionnaire three weeks after deployment.
This work makes three contributions:

\begin{itemize}
    \item We identify and operationalize three participant-informed BCT supports for pre-meal photo capture, namely Prompts/cues, Social support, and Focus on past success. We document how literature-informed candidate selection, participant evaluation, and considerations for smartphone implementation shaped these support conditions.
    
    \item We provide a longitudinal empirical account of pre-meal photo capture across support, withdrawal, reinstatement, and follow-up. The study traces capture adherence, perceived habit, and support use, showing broadly maintained capture and early subjective improvements, while support use, needs, and meanings changed over time.

    \item We derive design implications for supporting pre-meal photo capture as an everyday habit. Our findings suggest matching BCT support to changing difficulties in remembering, acting, and maintaining engagement, while preserving support availability and enabling personal records, tools, and relationships to continue supporting capture.
\end{itemize}




%% file: section/2_related_works_background.tex
\section{Related Work}

\subsection{Food Journals and Pre-Meal Photo Capture}

Food journals help people review meals~\cite{cordeiro2015rethinking}, identify and reflect on dietary patterns~\cite{blair2018onenote}, and plan dietary changes aligned with personal goals~\cite{chung2019identifying}. They also support collaborative review with nutrition professionals~\cite{chung2016boundary}, communication within families~\cite{lukoff2018tablechat}, and sharing records with peers~\cite{baumer2012prescriptive}.

Entering and organizing food information in conventional manual journals can be difficult and time-consuming, and users may omit foods from their records~\cite{chen2019use,bakar2025snappymeal}. 
These demands can also act as negative nudges, leading users to favor foods that are easier to log even when those choices conflict with their health goals~\cite{cordeiro2015barriers}. 
To reduce the effort required for active recording, Foody Talk explores how conversational interfaces can support records tailored to personal goals and everyday situations~\cite{silva2025foody}, while FoodScrap captures food descriptions and eating contexts through speech~\cite{luo2021foodscrap}. These approaches seek to reduce recording burden through more flexible input and interaction.
Wearable sensing can also ease recording by helping users remember when to log. FitNibble, for example, detects eating with eyeglass-mounted sensors and prompts food entry, reducing missed entries during field deployment~\cite{bedri2022fitnibble}. Continuous sensing may nevertheless raise concerns about physical burden and privacy, while wearing and managing devices can introduce social and interaction burdens~\cite{lu2022understanding,wilde2007concept}.

Smartphone-based photo journals provide a lightweight way to record meals without additional dedicated sensing hardware. They preserve the appearance of food and some contextual cues without requiring item-by-item entry of ingredients or nutritional information during capture~\cite{cordeiro2015rethinking}. Taking photos can encourage awareness during recording~\cite{zepeda2008think}, while subsequent review supports reflection on eating patterns in relation to personal goals and understandings of healthy eating~\cite{cordeiro2015rethinking,chung2019identifying}.

Making capture easier, however, does not resolve all difficulties associated with recording in everyday life. Even with lightweight photo journals, users may forget to record or feel uncomfortable photographing meals in social settings~\cite{cordeiro2015rethinking}. Beyond these practical barriers, diet-related systems often emphasize specific target behaviors while giving insufficient attention to personal meaning-making and the lived experience of change~\cite{rapp2025promoting}. As \citet{rooksby2014personal} note, people may turn to tracking technologies to pursue changes they already wish to make.
We therefore investigate how pre-meal photo capture becomes habitual. We define it as the simple act of photographing a main meal before the first bite and consider both repeated performance and people's experiences of the practice in everyday meal contexts.

\subsection{Behavior Change Techniques in HCI}

BCTs are identifiable, replicable, and irreducible intervention components. BCTTv1 groups 93 techniques into 16 categories, providing a common vocabulary for describing and comparing interventions~\cite{michie2013behavior}. They specify intervention content, whereas behavioral theories explain how change may occur~\cite{michie2013behavior,hekler2013mind}.

HCI research has used several BCTs, including Goal setting, Self-monitoring, Feedback, and Prompts/cues~\cite{wang2019health,zhu2025systematic}. Design cards map BCTs to stages of behavior change to guide target-specific selection~\cite{konstanti2022behavior}. These classifications and tools organize the design space, but do not determine which techniques suit a specific target behavior or how they should be implemented.
Recent systems have implemented behavior change support through contextual notifications~\cite{chen2024investigating}, action plans combining personal data, implementation intentions, and mental contrasting~\cite{sefidgar2024improving}, and reflective dialogue for mindfulness practice~\cite{kumar2025large}. However, some studies cite BCTs to justify design decisions without fully applying them to strategy selection and feature design~\cite{zhu2025systematic}. When multiple support features are combined and evaluated only at the system level, the effects and applicability of individual BCTs become difficult to distinguish~\cite{klasnja2011evaluate}.

Understanding the acceptability and effects of behavioral support in everyday social life also requires attention to interaction demands and daily contexts, in addition to theoretical and feature-level descriptions~\cite{hekler2013mind,consolvo2009theory}.
Evaluating support for habit formation requires attention to automaticity, sustained repetition, and integration into daily life, in addition to short-term behavioral performance. Repetition in consistent contexts can promote automaticity, but the rate and trajectory of development vary across individuals~\cite{lally2010habits}. In a smartphone recording task, reminders supported repetition without clear gains in automaticity~\cite{stawarz2015beyond}. Behavioral performance and habit formation are therefore distinct goals.
Sustained change also involves valuing the target behavior and maintaining motivation, not just system engagement~\cite{alberts2026designing}. Longitudinal research can reveal how behavior changes during continued use. In a four-week exercise intervention, 
visual self-modeling maintained relatively high performance, but improvements slowed over time, showing that performance level and rate of change are distinct~\cite{he2026does}.  
Understanding these changes requires considering personal goals, life circumstances, and subjective experience~\cite{rapp2023exploring}. Evaluation should also examine how specific support strategies work and how users experience them~\cite{klasnja2011evaluate}.

We therefore examine BCT support tailored to pre-meal photo capture through formative selection and longitudinal evaluation of behavioral trajectories, experiences of habit formation, and contextual influences







%% file: section/3_formative_selection.tex
\section{Formative Selection of BCTs}


Supporting behavior change in everyday life requires attention to users' routines and social contexts~\cite{consolvo2009theory}. We therefore conducted a two-stage formative process to determine \textit{which candidate BCTs participants considered potentially useful and acceptable for pre-meal photo capture, how these BCTs could be implemented using smartphones, and what considerations warranted their longitudinal evaluation}. Screening informed by the HCI and behavioral science literature yielded an initial candidate set. Two workshops then elicited participants' evaluations and implementation ideas to guide final BCT selection and support design.

\begin{table*}[t]
    \centering
    \small
    \setlength{\tabcolsep}{4pt}
    \renewcommand{\arraystretch}{1.08}
    \begin{tabularx}{\textwidth}{
        @{}
        >{\raggedright\arraybackslash}p{0.27\textwidth}
        >{\centering\arraybackslash}p{0.09\textwidth}
        >{\raggedright\arraybackslash}X
        @{}
    }
        \toprule
        BCT &
        Sources &
        Preliminary inclusion consideration \\
        \midrule
        \midrule
        7.1 Prompts/cues &
        EH, RH, BS &
        Directly targets forgetting and timing by cueing capture near
        meal onset, while leaving cue timing and delivery open for
        workshop exploration. \\

        2.2 Feedback on behavior &
        EH, RH, BS &
        Makes recent capture behavior visible through neutral summaries
        that can remain distinct from goal comparison, praise, and
        outcome feedback. \\

        1.4 Action planning &
        EH, RH, BS &
        Links capture to a concrete cue within an existing meal routine,
        offering a context-specific approach to supporting
        self-initiated action. \\

        3.1 Social support (unspecified) &
        EH, RH, BS &
        Offers behavior-directed encouragement from a trusted person or
        facilitator, while allowing the workshop to examine provider,
        interaction, and social-burden considerations. \\

        1.6 Discrepancy between current behavior and goal &
        EH, RH &
        Could make gaps between actual capture and an existing goal or
        plan salient; its dependence on prior goals and possible overlap
        with feedback warranted further examination. \\

        6.2 Social comparison &
        EH, BS &
        Provides a distinct social reference by relating a person's
        recording behavior to others' performance, warranting evaluation
        of its acceptability and possible social costs. \\

        1.2 Problem solving &
        BS &
        Directly addresses recurring capture barriers by combining
        analysis of difficult situations with concrete coping
        strategies. \\

        8.7 Graded tasks &
        BS &
        Offers a gradual progression from easier capture tasks to broader
        meal coverage, while raising questions about burden and changing
        task demands. \\

        10.4 Social reward &
        TD &
        Provides a phone-deliverable form of contingent positive reinforcement after successful capture that can remain distinct from behavior feedback, goal comparison, and advance incentives.
        \\

        15.3 Focus on past success &
        TD &
        Uses previous successful captures as concrete material for reflection, offering a support direction distinct from behavior counts, praise, and future planning.\\

        \bottomrule
    \end{tabularx}
    \caption{The ten BCTs retained for workshop discussion and design. Inclusion considerations reflect preliminary, study-specific judgments. EH: HCI research through 2018 summarized by \citet{wang2019health}. RH: HCI research since 2019 reviewed in this study. BS: the public-health and behavioral-science corpus. TD: additional candidates proposed through team discussion.}
    \label{tab:candidate_bcts}
\end{table*}

\subsection{Candidate Identification}

To identify a manageable set of BCTs for workshop discussion and design, we applied a staged, literature-informed selection process to the 93 techniques in BCTTv1~\cite{michie2013behavior,marques2021theory,garnett2019development}. Using search terms adapted from \citet{wang2019health}, we searched the ACM Digital Library for health behavior change articles published in CHI, DIS, UIST, IMWUT, and CSCW since 2019. After deduplication, title and abstract screening, and full-text review, we included 64 papers that reported the implementation and evaluation of participant-facing health behavior change interventions. We coded the intervention content using BCTTv1 and identified 47 distinct techniques. 
Appendix~\ref{app:literature_screening} details the search, screening, and coding procedures and reports the screening flow following PRISMA~\cite{page2021prisma}.
We then integrated three literature-based candidate pools: 10 BCTs summarized by \citet{wang2019health} from HCI research through 2018, the 10 BCTs most frequently identified in our review, and 16 BCTs identified in the public health and behavioral science corpus indexed on the BCTTv1 website\footnote{\url{https://www.bct-taxonomy.com/}}. After deduplication, these pools yielded 17 techniques.
Through team discussion, we added two lower-frequency techniques that offered distinct forms of support for pre-meal photo capture and could be delivered through smartphones. Social reward provides contingent positive reinforcement after successful capture, and Focus on past success draws on previous successful captures to support self-belief. This produced a pool of 19 BCTs.

Using the BCT definitions and possible implementations, the research team assessed whether each candidate BCT offered target-specific value beyond the shared study procedures, could be implemented distinctly from other BCTs and the study task, and was compatible with the smartphone-only scope.
These considerations informed the preliminary selection of ten candidates for workshop discussion. Table~\ref{tab:candidate_bcts} shows the source of each candidate and the considerations supporting inclusion. Appendix~\ref{app:bct_selection} documents the composition of the literature-based pools and reasons for excluding the remaining BCTs.

\subsection{Workshops for BCT Selection and Support Design}

We conducted a formative workshop study comprising two parallel sessions and a subsequent individual reflection phase. We examined how participants evaluated candidate BCTs and used them to develop smartphone-based support concepts for pre-meal photo capture.

\subsubsection{Participants}

We recruited 10 participants (2 male, 8 female; $M=24.8$, $SD=3.1$) through social media and campus communication platforms. Participants were eligible if they considered developing a habit of pre-meal photo capture personally meaningful. None reported substantial prior familiarity with BCTs, although five had heard of them.
We assigned participants to two workshop sessions based on their recent self-reported capture rates, as shown in Table~\ref{tab:workshop_participants}. Both sessions included participants with different levels of engagement and practical experience with the target behavior. The study was approved by the Institutional Review Board (IRB) of our university. Participants received USD 10 per hour.

\begin{table}[b]
    \centering
    \begin{tabular}{ccccc}
        \toprule
        ID & Gender & Age & Past-Week Capture & Longest Continuous Recording \\
        \midrule
        \midrule
        \multicolumn{5}{l}{Workshop 1} \\
        P1 & M & 21 & 9 / 9   & 1--3 months \\
        P2 & F & 28 & 6 / 13  & $\leq$3 days \\
        P3 & F & 23 & 3 / 14  & 1--3 months \\
        P4 & F & 23 & 12 / 16 & 4--7 days \\
        P5 & M & 25 & 11 / 14 & 4--7 days \\
        \midrule
        \multicolumn{5}{l}{Workshop 2} \\
        P6 & F & 24 & 3 / 14  & 1--3 months \\
        P7 & F & 21 & 14 / 19 & $>$3 months \\
        P8 & F & 26 & 5 / 21  & $\leq$3 days \\
        P9  & F & 31 & 7 / 10  & 4--7 days \\
        P10 & F & 26 & 17 / 18 & $>$3 months \\
        \bottomrule
    \end{tabular}
    \caption{Participant characteristics and self-reported baseline recording practices. Past-week capture is reported as the number of main meals photographed before eating divided by the total number of main meals during the preceding week.}
    \label{tab:workshop_participants}
\end{table}

\subsubsection{Workshop Materials}

Drawing on probe-based design research and co-design workshops~\cite{gaver1999design,chen2025vifeed}, we prepared four sets of generative materials to support reflection and BCT-specific ideation.

\paragraph{Context-reflection cards.} 

To help participants revisit prior attempts at pre-meal photo capture~\cite{kocielnik2018reflection}, the cards presented routine, time-pressured, and social meals as three recurring contexts. Participants rated each context in terms of frequency, importance, and difficulty.


\paragraph{BCT concept cards.} 

Following work that translates behavior-change knowledge into design materials~\cite{konstanti2022behavior,caraban2020nudge}, we created one card for each candidate BCT. Each card provided an accessible description, clarified what the standard definition included and excluded, and illustrated a minimal smartphone-based implementation for pre-meal photo capture~\cite{michie2013behavior}. The examples were intended to support interpretation without prescribing final designs.


\paragraph{Paper-based design probe.} 

An A3 ideation sheet prompted participants to consider five dimensions: \textbf{HOW} the support would operate, \textbf{WHEN} it should intervene, \textbf{WHERE} it would apply across meal contexts, how the support could be delivered through a \textbf{PHONE}, and how its role might change \textbf{OVER TIME}.


\paragraph{Colored sticky notes.} 
Participants used colored notes first to record their initial positive and negative judgments and the reasons for them, and later to indicate whether each BCT should be developed and evaluated further.


\begin{figure}[b]
    \centering
    \includegraphics[width=\linewidth]{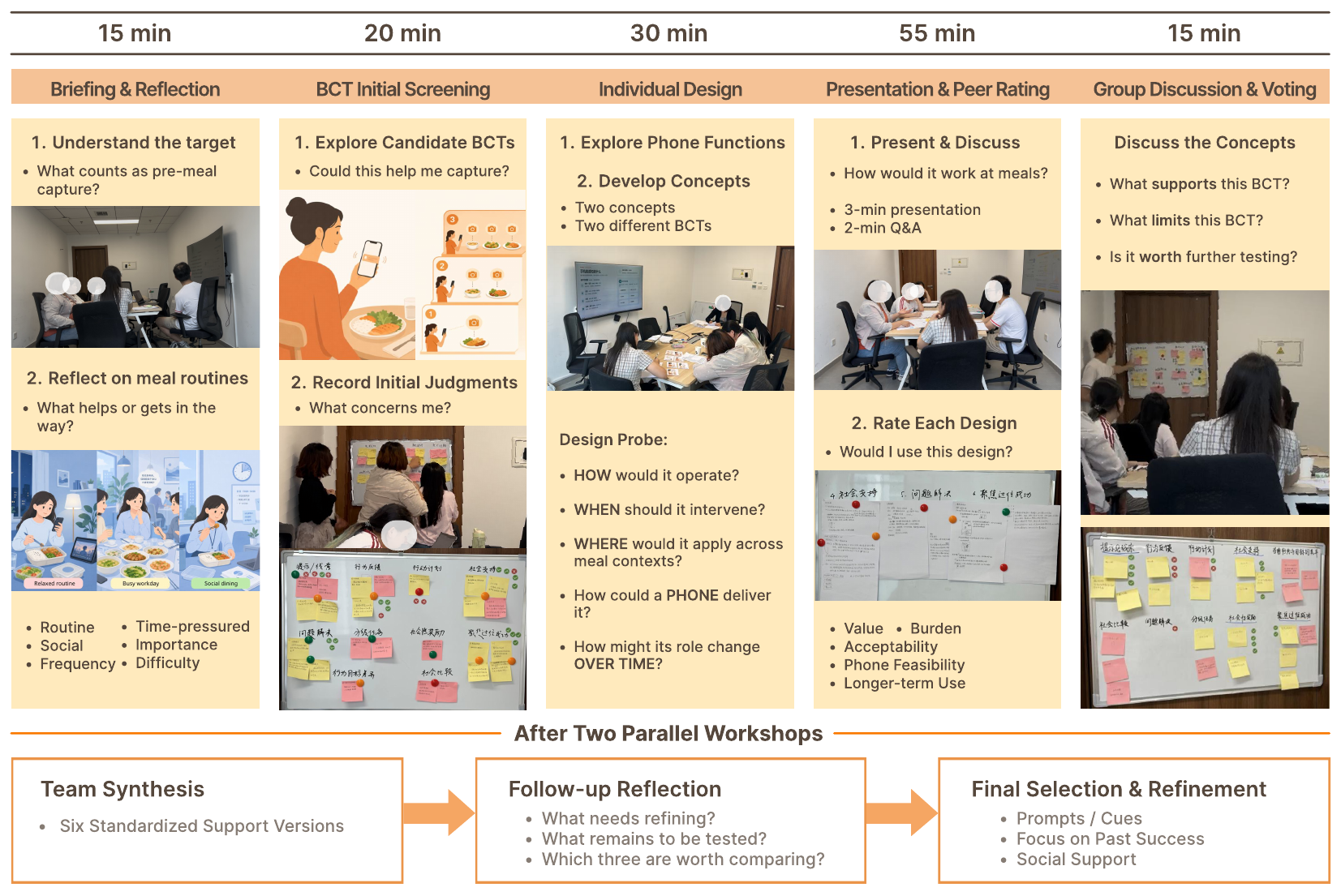}
    \caption{Overview of the formative BCT selection and support-design process. Two parallel workshops progressed from target-behavior reflection and initial BCT screening to individual smartphone concept design, peer rating, and BCT-level discussion. The research team then synthesized six standardized support versions for individual follow-up reflection, leading to the final selection and refinement of Prompts/cues, Focus on past success, and Social support.}
    \label{fig:workshop_format}
\end{figure}

\subsubsection{Workshop Format}

We conducted two parallel workshops using the five-stage process summarized in Figure~\ref{fig:workshop_format}, after which participants completed an individual reflection phase. Each workshop was facilitated by one researcher familiar with BCTs and one co-facilitator with design experience.
Participants first reflected on meal contexts and screened the candidate BCTs, after which each participant developed two smartphone concepts based on different BCTs. They presented and peer-rated these concepts before discussing and voting at the BCT level. Based on discussion and concepts from two workshops, the research team synthesized a standardized version of support for the six retained BCTs. Then, participants independently evaluated these versions, proposed refinements, identified questions for longitudinal evaluation, and nominated three BCTs for the field study. Detailed stage-level procedures and evaluation instruments are provided in Appendix~\ref{app:workshop_evaluation}.

\subsubsection{Data Analysis and Selection Process}

We analyzed quantitative data separately by evaluation target. For each concept, we calculated medians and interquartile ranges for peer ratings with and without self-ratings by the design authors. We combined the three context-fit items using each respondent's normalized context-importance ratings as weights and analyzed all other items separately. We also summarized screening results for each workshop, BCT-level Yes/No judgments, and follow-up responses.

We interpreted these results alongside a BCT-centered qualitative analysis. Following the matrix organization of the Framework Method~\cite{gale2013using}, we organized sticky notes, design artifacts, workshop discussions, and follow-up responses by BCT. The matrix captured anticipated support pathways, contextual conditions, burdens and risks, implementation requirements, dependencies on or overlap with other BCTs, and expectations during continued use or after support withdrawal. We compared these qualitative accounts with the quantitative results to identify agreement, disagreement, and implementation-specific differences.


Selection proceeded iteratively. BCTs excluded during initial screening in both workshops were not considered further. Based on the peer ratings, we retained concepts ranked in the top 50\% of the distribution as evidence of potentially promising implementations of the corresponding BCTs.
We then reviewed these concepts, the related discussions, and the BCT-level Yes/No judgments collected at the end of each workshop. BCTs for which Yes votes exceeded No votes in at least one workshop were retained for follow-up reflection.
In selecting the final three BCTs, we considered follow-up ratings, advancement judgments, participant nominations, and qualitative reasoning. BCT-level analytic memos supported candidate selection and refinement of the support designs. They also documented the evidence underlying the questions taken forward into the longitudinal study.

\subsection{Selection Outcomes and Design Implications}
\label{sec:formative_selection}

We report how the formative analysis narrowed ten candidate BCTs to three for longitudinal evaluation and identify the design implications carried forward.



\subsubsection{Selection Trajectory}




Both parallel workshops began with the same ten candidate BCTs. Initial screening retained eight candidates in Workshop~1 and six in Workshop~2. Discrepancy between current behavior and goal and Social comparison were excluded in both workshops because participants offered no positive rationale for retaining either technique. For Discrepancy between current behavior and goal, P5 and P10 warned that \textit{"visible gaps or unmet goals could undermine motivation, cause frustration, and prompt abandonment"}. For Social comparison, P1 and P2 saw little need to compare their recording behavior with that of others, while P8 worried that comparison would \textit{"shift attention from one's own goal to others' behavior"}.


Six BCTs then advanced to the follow-up reflection: Prompts/cues, Feedback on behavior, Social support, Problem solving, Social reward, and Focus on past success. BCT-level judgments sometimes differed across workshops. For example, Feedback on behavior received three No judgments in Workshop~1 but three Yes judgments and one No judgment in Workshop~2. We therefore gathered additional perspectives during the follow-up reflection and narrowed the set to Prompts/cues, Focus on past success, and Social support. Table~\ref{tab:bct-selection} summarizes the complete selection trajectory and the main participant accounts associated with each candidate.

\begin{table*}[t]
\centering
\footnotesize
\setlength{\tabcolsep}{3pt}
\renewcommand{\arraystretch}{1.05}
\renewcommand{\tabularxcolumn}[1]{m{#1}}
\begin{tabularx}{\textwidth}{@{}
  >{\raggedright\arraybackslash}m{0.120\textwidth}
  >{\centering\arraybackslash}m{0.027\textwidth}
  >{\centering\arraybackslash}m{0.027\textwidth}
  >{\centering\arraybackslash}m{0.075\textwidth}
  >{\centering\arraybackslash}m{\dimexpr0.040\textwidth-\tabcolsep\relax}
  >{\centering\arraybackslash}m{\dimexpr0.040\textwidth-\tabcolsep\relax}
  >{\hsize=0.88\hsize\centering\arraybackslash}X
  >{\hsize=1.12\hsize\centering\arraybackslash}X
  >{\centering\arraybackslash}m{0.072\textwidth}@{}}
\toprule
\textbf{BCT} &
\multicolumn{5}{c}{\textbf{Screening and design evidence}} &
\multicolumn{2}{c}{\textbf{Participant judgments}} &
\shortstack{\textbf{Final}\\\textbf{selection}} \\
\cmidrule(lr){2-6}\cmidrule(lr){7-8}
& \multicolumn{2}{c}{\shortstack{Initial\\screening}}
& \shortstack{Top-half\\concept}
& \multicolumn{2}{c}{\shortstack{BCT vote\\Yes $>$ No}}
& \textbf{Perceived value}
& \shortstack{\textbf{Concerns / conditions}}
& \\
& {\scriptsize W1} & {\scriptsize W2} & {\scriptsize Incl./excl.} & {\scriptsize W1} & {\scriptsize W2} & & & \\
\midrule
\midrule








Social support &
$\checkmark$ &
$\checkmark$ &
$\checkmark$ / $\checkmark$ &
$\checkmark$ &
$\checkmark$ &
Sharing with trusted others could motivate subsequent photo recording~(P4). &
Supporter availability and burden~(P1, P9).\par Reliance on others' reminders~(P6).\par Reciprocal sharing did not require replies~(P7). &
$\checkmark$ \\

\addlinespace[3pt]









Focus on past success &
$\checkmark$ &
$\checkmark$ &
$\checkmark$ / $\checkmark$ &
$\checkmark$ &
$\checkmark$ &
Personal successes could build confidence to continue recording~(P9).\par Review could make persistence visible~(P7). &
Success could feel trivial or discouraging after a lapse~(P3).\par Self-initiated reviews might not occur~(P7). &
$\checkmark$ \\

\addlinespace[3pt]








Prompts/cues &
$\checkmark$ &
$\checkmark$ &
$\checkmark$ / $\checkmark$ &
$\checkmark$ &
$\times$ &
Could bring photo capture to mind before eating~(P6, P7). &
Fixed alarms could miss irregular meals~(P10).\par Alarms could disrupt others~(P8).\par Continued reliance raised withdrawal concerns~(P1). &
$\checkmark$ \\

\midrule







Feedback on behavior &
$\checkmark$ &
$\checkmark$ &
$\checkmark$ / $\checkmark$ &
$\times$ &
$\checkmark$ &
Recording summaries could reveal personal patterns~(P1, P9).\par They could also strengthen confidence~(P8). &
Basic summaries could feel too thin~(P6).\par Counts alone might offer little motivation and be ignored~(P3). &
$\times$ \\

\addlinespace[3pt]









Problem solving &
$\checkmark$ &
$\checkmark$ &
$\times$ / $\checkmark$ &
$\times$ &
$\checkmark$ &
Could suggest ways to address barriers to pre-meal capture~(P3, P6). &
Advice depended on context~(P3, P5).\par Users might not recognize a problem or seek strategies~(P9, P10). &
$\times$ \\

\addlinespace[3pt]








Social reward &
$\checkmark$ &
$\times$ &
$\checkmark$ / $\checkmark$ &
$\checkmark$ &
\textemdash &
Praise could encourage continued recording~(P1, P3). &
Praise could have short-lived effects~(P6).\par Repetition could reduce its impact~(P1).\par Rewards would not appear when capture was forgotten~(P3). &
$\times$ \\

\midrule








Action planning &
$\checkmark$ &
$\checkmark$ &
$\checkmark$ / $\checkmark$ &
$\times$ &
$\times$ &
Could clarify intended actions~(P6).&
Fixed timing could miss delayed meals~(P2).\par Plans alone seemed limited without problem solving~(P8). &
$\times$ \\

\addlinespace[3pt]









Graded tasks &
$\checkmark$ &
$\times$ &
$\checkmark$ / $\checkmark$ &
$\times$ &
\textemdash &
Gradually extend recording from meaningful meals to everyday meals~(P4). &
Rules added effort~(P9).\par Stalled progression could discourage~(P7, P9).\par One design relied on past-success review~(P4). &
$\times$ \\

\midrule







Discrepancy between current behavior and goal &
$\times$ &
$\times$ &
\textemdash &
\textemdash &
\textemdash &
\textemdash &
Questioned fit with routine recording~(P4).\par Frustration over unmet goals~(P10).\par Discouragement from large gaps~(P5). &
$\times$ \\

\addlinespace[3pt]









Social comparison &
$\times$ &
$\times$ &
\textemdash &
\textemdash &
\textemdash &
\textemdash &
Comparison seemed unnecessary for a personal habit~(P1, P2).\par It could shift attention to others~(P8).\par It could create pressure or shame~(P4, P7). &
$\times$ \\

\bottomrule
\end{tabularx}
\caption{Candidate progression and participant assessments of support for pre-meal photo capture. W1 and W2 involved different participants. $\checkmark$/$\times$ indicate whether the condition named in each column was met. \textemdash{} denotes a stage not reached. Top-half concept indicates at least one design among the top 10 of 20, including/excluding author self-ratings (Incl./excl.). BCTs receiving more Yes than No votes in at least one workshop's final vote entered follow-up reflection. Final selection denotes inclusion in the longitudinal study. Participant IDs identify sources of summarized expectations.}
\label{tab:bct-selection}
\end{table*}

\subsubsection{Selection Rationale Across Evaluation Levels}

Because peer ratings assessed specific concepts, decisions about BCT advancement relied mainly on converging evidence across evaluation levels. Action planning and Graded tasks each had at least one concept in the upper half of the peer-rating distribution. Action planning also received positive concept-level judgments, but neither BCT advanced. P8 considered Action planning alone unlikely to influence behavior and suggested combining it with Problem solving. P4's Graded tasks concept relied on reviewing prior achievements and therefore overlapped with Focus on past success. P7 and P9 also anticipated pressure from unfinished levels and the effort required to manage task rules.

These dependencies and demands limited the value of both BCTs as distinct support options despite favorable ratings for specific concepts.
Across the six shortlisted candidates, advancement also depended on whether support could be provided at useful moments in everyday meal routines and whether users could initiate it with manageable effort. These conditions helped explain differences across evaluation stages and informed the final prioritization.

\paragraph{From six candidates to three.}





In the follow-up reflection, all six standardized support versions received high smartphone-feasibility ratings, with medians of 6--7. The final selection therefore prioritized candidates that could support an informative longitudinal comparison beyond basic feasibility. Prompts/cues received 8 nominations, Social support 7, and Focus on past success 5. All three counts exceeded those for the remaining candidates.
The three candidates not prioritized had plausible support pathways but also important limitations. Feedback on behavior could make recording patterns visible, but participants differed on whether basic summaries would support subsequent capture. P3 considered \textit{"count summaries easy to ignore and unlikely to motivate capture"}. We therefore prioritized reflection on a specific personal success while retaining Feedback on behavior as a feasible alternative. Problem solving could address concrete barriers but required users to recognize a problem, provide sufficient context, and seek or adopt a strategy. P10 noted, \textit{"I might not perceive difficulty with photo capture as a problem and therefore might not use the support"}. Social reward could encourage users after successful capture but could not intervene when capture was missed and might lose impact through repetition. These limitations, combined with lower nomination counts, reduced the priority of these candidates for longitudinal comparison.

The three selected BCTs represented distinct forms of external cueing before meals, interpersonal support, and off-meal self-reflection. Their differences in intervention timing, required user initiation, potential burden, and reliance on continued support made them informative candidates for longitudinal comparison.


\subsubsection{Selected BCTs and Design Implications}
\label{sec:bct_design_implication}
For each selected BCT, we summarize the participant accounts that supported its selection, the design requirements derived from these accounts, and the considerations carried into the longitudinal study. These findings informed the subsequent operationalization of these BCTs as smartphone-based support conditions.

\paragraph{Prompts/cues.}


Prompts/cues were valued for bringing capture to mind before eating (P6, P7), but fixed alarms raised concerns about irregular mealtimes (P10) and disruption to others (P8).  
We therefore specified low-frequency, user-configurable cues delivered near personal mealtimes, using less disruptive modalities and providing direct access to photo capture. These requirements preserve the immediate cueing function while limiting interaction and attentional burden. These considerations informed our longitudinal analysis of prompt timing, burden, reliance, and changes in capture after support withdrawal.

\paragraph{Focus on past success.}


We operationalized success as photographing a main meal before the first bite. Participants valued using personally meaningful successes as references for continued recording (P9) and seeing evidence that they had sustained the practice (P7). However, P3 questioned whether an ordinary capture would feel meaningful, and P8 noted the challenge of starting without prior records. These accounts led us to specify two complementary forms of review, one based on successful captures across days and the other on personally meaningful successes selected by the participant. To address concern that repeated material might become familiar (P2), we included options to refresh the review content. The design also provided a starting point for users without prior app records.
We kept the presentation neutral by excluding completion statistics, praise, problem-solving advice, and future planning.
These considerations informed our longitudinal analysis of engagement with the reflection, its relationship to subsequent capture, and changes in perceived value during continued use and after withdrawal.

\paragraph{Social support.}


Participants associated sharing and receiving responses with motivation to continue recording. Although P6 envisioned ongoing reminders from a supporter, P7 preferred voluntary sharing without requiring a response to every meal. P9 and P1 also raised concerns about supporter availability and social burden. We therefore specified a bounded, user-initiated arrangement that preserved control over sharing, permits brief, nonjudgmental responses, and did not require continuous supervision or a reply to every meal. We also separated Social support from mealtime reminders and contingent praise to preserve the conceptual boundary of the BCT. In the longitudinal study, we examined how participants used this interaction, whether it created burden, pressure, or response expectations, and how capture changed as support decreased or was withdrawn.

%% file: section/4_system_design.tex
\section{Mobile System Design}



\begin{figure}[t]
    \centering
    \includegraphics[width=\linewidth]{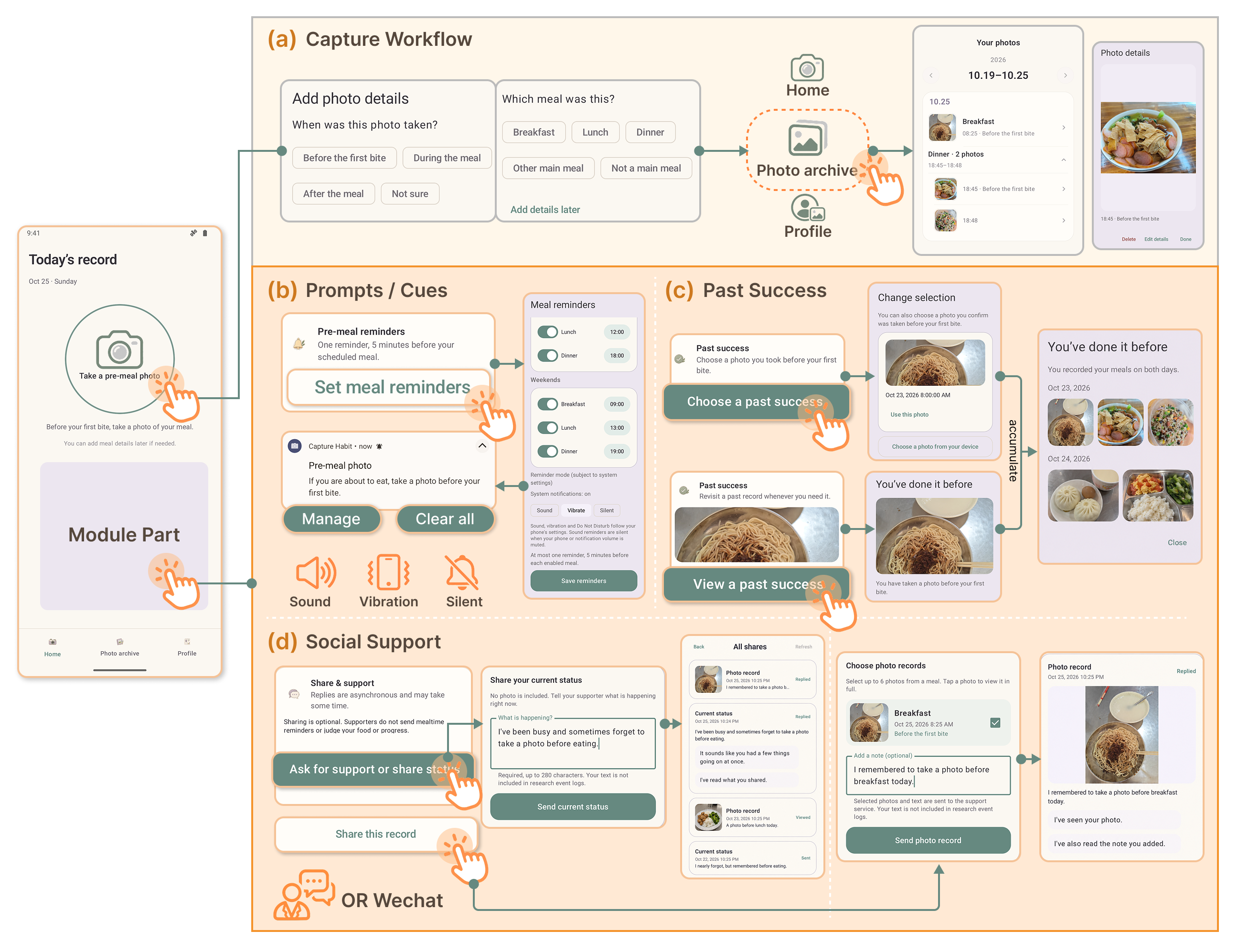}
    \caption{\textit{MealCapture}'s shared workflow and BCT-based support modules. (a) Photo capture, meal annotation, and archive access; (b) configurable meal-time reminders; (c) selecting and revisiting personal capture successes; and (d) participant-initiated sharing with a researcher via WeChat or an in-app chatbot. All conditions shared the same capture and archive functions, with each participant receiving one assigned support module.}
    \label{fig:system_workflow}
\end{figure}

To operationalize the three selected BCTs for a longitudinal study, we developed \textit{MealCapture}, an Android app for pre-meal photo capture. All three versions shared the same capture, classification, archive, and navigation workflows. Each version added a support module based on the design implications identified in Section~\ref{sec:bct_design_implication}. We kept each module focused on its corresponding BCT and pre-meal photo capture to maintain a comparable functional scope while preserving differences in support timing and required user initiation. Figure~\ref{fig:system_workflow} summarizes the shared workflow and the support module for each condition.

In all versions, the home screen provided the same entry point for photographing a main meal before the first bite. Participants could label capture timing and meal type immediately after saving a photo or later in the app archive. To reduce repeated labeling, subsequent photos within a predefined time window could inherit the meal type from a previously classified record.

\subsection{Prompts/cues}

To accommodate personal meal routines and interruption preferences, participants could set breakfast, lunch, and dinner times separately for weekdays and weekends, enable or disable each meal slot, and choose among sound, vibration, and visual-only notification modes.
To limit repeated interruptions, \textit{MealCapture} scheduled a single neutral reminder five minutes before each enabled meal time, prompting participants to capture the pre-meal photo. Opening the notification took participants to the home screen, where they could initiate capture through the shared camera entry point.

\subsection{Focus on past success}


This module displayed a preview of selected photos that participants could open to revisit successful pre-meal captures.
Participants could start with either an app record marked as a main meal photo captured before the first bite or an earlier photo they regarded as a successful pre-meal capture. This provided a starting point before sufficient app records had accumulated.
To support reflection on continued recording, \textit{MealCapture} organized review material from a single day or several consecutive days. For automatic updates, the module prioritized available material from captures made on consecutive days, followed by material from a single day and then the initial photo.
Participants could retain the current selection, choose different material, or resume automatic updates. These controls allowed personally meaningful experiences to remain available for review. The module grouped photos by date and presented brief, neutral, factual descriptions.

\subsection{Social support}



To accommodate preferences for support providers and disclosure, and to account for differences in how human and chatbot support may be experienced~\cite{meng2023mediated}, the module offered support from either a study researcher or an in-app LLM chatbot. 
Participants could select pre-meal photos in the app and share them with the researcher through WeChat, optionally adding a short message. Alternatively, they could send multiple photos with an optional note to the chatbot through the module interface, or send text alone to describe their current situation or support needs. Both routes provided brief, nonjudgmental responses without continuous monitoring or an expectation of a reply to every meal record.
The researcher responded upon viewing incoming messages.
Responses acknowledged participants' recording experiences and offered support without meal reminders, action plans, or praise contingent on successful capture.

To vary the wording within this scope, the research team prepared a pool of 100 conversational chatbot responses, each containing one to three sentences. For a small proportion of submissions, the interface displayed a simulated read receipt instead of a textual reply. For the remaining submissions, the chatbot selected an appropriate response from the pool. Replies appeared immediately or after a delay of 5--60 minutes, according to predefined probabilities. Submitted messages and subsequent responses remained available in the chat history.

\subsection{Implementation}

\textit{MealCapture} was implemented as a native Android application in Kotlin with Jetpack Compose. The application invoked the device camera through the Android Activity Result API and stored photographs in app-private storage. Capture metadata, condition states, and timestamped interaction events were stored locally using Room.
The Prompts/cues module used Android AlarmManager to schedule notifications based on participant-configured meal times.
The Focus on past success module used local rules to assemble and update review material from stored records.
For chatbot submissions requiring a reply, AndroidX WorkManager processed image and text requests asynchronously, and \textit{MealCapture} called Kimi K2.6 over HTTPS to select a response from the predefined pool.

%% file: section/5_longitudinal_study.tex
\section{Longitudinal Field Study}



To examine how the selected BCTs support habit formation in pre-meal photo capture, we conducted a five-week field study with three parallel groups. The study comprised two weeks of support, one week of withdrawal, and two weeks of reinstatement, allowing us to trace changes in capture adherence, experiences of habit formation, and the use and perceived value of support across everyday meal contexts.

\subsection{Participants}

\begin{table*}[t]
    \centering
    \footnotesize
    \setlength{\tabcolsep}{0.8pt}
    \renewcommand{\arraystretch}{0.96}
    \begin{tabular*}{\textwidth}{@{\extracolsep{\fill}}
        rccccccc@{\hspace{8pt}}
        rccccccc@{\hspace{8pt}}
        rccccccc@{}}
        \toprule
        \multicolumn{8}{c}{\textbf{PC: Prompts/cues}} &
        \multicolumn{8}{c}{\textbf{PS: Focus on past success}} &
        \multicolumn{8}{c}{\textbf{SS: Social support}} \\
        \cmidrule(r){1-8}\cmidrule(lr){9-16}\cmidrule(l){17-24}
        ID & Age & Sex & IC & SRBAI & R & N & Cov. &
        ID & Age & Sex & IC & SRBAI & R & N & Cov. &
        ID & Age & Sex & IC & SRBAI & R & N & Cov. \\
        \midrule
        \midrule
        PC1 & 20 & F & 6.00 & 2.25 & 2 & 4 & $<25\%$
        & PS1 & 20 & M & 6.50 & 6.00 & 4 & 3 & 25--50\%
        & SS1 & 22 & M & 6.50 & 4.50 & 5 & 3 & 25--50\% \\

        PC2 & 23 & F & 5.83 & 2.50 & 3 & 6 & $<25\%$
        & PS2 & 25 & F & 6.83 & 1.50 & 4 & 5 & $<25\%$
        & SS2 & 21 & F & 6.33 & 5.75 & 6 & 3 & 75--90\% \\

        PC3 & 19 & F & 7.00 & 5.00 & 3 & 4 & $<25\%$
        & PS3 & 22 & M & 6.67 & 6.00 & 7 & 3 & 25--50\%
        & SS3 & 25 & F & 7.00 & 6.75 & 6 & 3 & 50--75\% \\

        PC4 & 27 & F & 6.83 & 3.00 & 3 & 6 & 25--50\%
        & PS4 & 21 & F & 6.33 & 4.00 & 5 & 4 & 25--50\%
        & SS4 & 26 & F & 6.00 & 2.75 & 3 & 5 & Almost none \\

        PC5 & 40 & F & 6.67 & 5.50 & 7 & 3 & 25--50\%
        & PS5 & 20 & F & 6.17 & 4.25 & 4 & 5 & 50--75\%
        & SS5 & 34 & F & 5.50 & 2.50 & 3 & 5 & 25--50\% \\

        PC6 & 27 & F & 6.83 & 2.00 & 4 & 5 & 25--50\%
        & PS6 & 37 & F & 6.33 & 6.50 & 6 & 3 & 50--75\%
        & SS6 & 28 & F & 7.00 & 2.75 & 4 & 6 & 50--75\% \\

        PC7 & 20 & F & 5.50 & 2.25 & 3 & 6 & $<25\%$
        & PS7 & 33 & F & 7.00 & 1.25 & 1 & 7 & Almost none
        & SS7 & 25 & M & 6.00 & 2.25 & 2 & 6 & $<25\%$ \\

        PC8 & 25 & F & 6.83 & 1.50 & 1 & 5 & $<25\%$
        & PS8 & 30 & F & 6.50 & 2.00 & 2 & 4 & $<25\%$
        & SS8 & 48 & F & 6.67 & 3.50 & 4 & 4 & $<25\%$ \\

        PC9 & 21 & F & 5.67 & 4.00 & 5 & 4 & $<25\%$
        & PS9 & 26 & F & 6.33 & 4.25 & 4 & 6 & 50--75\%
        & SS9 & 25 & F & 6.67 & 3.00 & 3 & 5 & $<25\%$ \\

        PC10 & 25 & F & 7.00 & 6.75 & 7 & 1 & $<25\%$
        & PS10 & 27 & F & 6.67 & 3.50 & 3 & 6 & $<25\%$
        & SS10 & 25 & F & 6.17 & 3.00 & 4 & 5 & $<25\%$ \\

        PC11 & 24 & F & 6.00 & 2.50 & 2 & 5 & Almost none
        & PS11 & 27 & F & 5.83 & 4.25 & 5 & 3 & 50--75\%
        & SS11 & 25 & F & 6.83 & 5.50 & 5 & 4 & 25--50\% \\

        PC12 & 19 & M & 5.83 & 5.50 & 5 & 5 & 50--75\%
        & PS12 & 28 & F & 6.83 & 7.00 & 7 & 4 & 25--50\%
        & SS12 & 23 & F & 6.83 & 5.50 & 7 & 4 & 25--50\% \\

        PC13 & 19 & M & 6.83 & 6.00 & 6 & 6 & 50--75\%
        & PS13 & 19 & F & 5.67 & 4.00 & 4 & 4 & $<25\%$
        & SS13 & 24 & F & 5.67 & 5.25 & 6 & 5 & 25--50\% \\

        PC14 & 20 & F & 6.83 & 6.00 & 7 & 2 & 25--50\%
        & PS14 & 21 & F & 6.67 & 6.00 & 6 & 2 & 75--90\%
        & SS14 & 28 & F & 6.00 & 6.50 & 6 & 2 & 50--75\% \\

        PC15 & 30 & F & 6.50 & 6.50 & 6 & 7 & 75--90\%
        & PS15 & 23 & F & 5.83 & 3.25 & 2 & 6 & $<25\%$
        & SS15 & 22 & M & 6.83 & 4.75 & 5 & 2 & Almost none \\

        PC16 & 29 & F & 6.33 & 5.25 & 4 & 6 & 50--75\%
        & PS16 & 28 & F & 6.17 & 3.25 & 2 & 6 & Almost none
        & SS16 & 26 & F & 5.83 & 2.50 & 3 & 5 & $<25\%$ \\

        PC17 & 23 & F & 6.33 & 5.00 & 5 & 3 & 25--50\%
        & PS17 & 21 & F & 6.83 & 4.50 & 5 & 5 & 25--50\%
        & SS17 & 24 & M & 6.33 & 3.75 & 5 & 6 & 25--50\% \\

        PC18 & 26 & M & 6.00 & 2.75 & 3 & 5 & $<25\%$
        & PS18 & 24 & F & 6.67 & 2.50 & 4 & 5 & $<25\%$
        & SS18 & 28 & F & 6.83 & 2.25 & 2 & 5 & 25--50\% \\

        \cmidrule(r){1-8}\cmidrule(lr){9-16}\cmidrule(l){17-24}

        Mean & 24.28 & -- & 6.38 & 4.13 & 4.22 & 4.61 & 31.1\%
        & Mean & 25.11 & -- & 6.44 & 4.11 & 4.17 & 4.50 & 33.3\%
        & Mean & 26.61 & -- & 6.39 & 4.04 & 4.39 & 4.33 & 33.3\% \\
        
        SD & 5.28 & -- & 0.50 & 1.77 & 1.86 & 1.58 & 23.5\%
        & SD & 4.90 & -- & 0.38 & 1.70 & 1.72 & 1.38 & 24.7\%
        & SD & 6.10 & -- & 0.47 & 1.53 & 1.50 & 1.28 & 23.2\% \\

        \bottomrule
    \end{tabular*}
    \caption{Participant characteristics and recording measures at baseline,
    with individual values and group means and sample SDs.
    F: female; M: male.
    IC: recording intention;
    SRBAI: capture automaticity;
    R: spontaneous remembering;
    N: need for deliberate self-reminding.
    Cov.: recent pre-meal capture proportion, self-reported for the
    preceding two weeks. Individual rows retain the original response
    categories; summary values are calculated from numerical estimates
    assigned to these categories.}
    \label{tab:participant_baseline}
    \label{tab:individual_participant_baseline}
\end{table*}

We recruited participants through university communication platforms and social media, using snowball sampling~\cite{goodman1961snowball}. Participants were eligible only if they wanted to establish a habit of pre-meal photo capture and valued saving and reviewing meal photos even without downstream functionality. We enrolled 54 participants (45 female, 9 male; ages 19--48 years, $M = 25.33$, $SD = 5.43$). We assigned 18 participants to each condition, including Prompts/cues (PC), Focus on past success (PS), and Social support (SS). All enrolled participants scored at least 5 out of 7 on the recording intention. Individuals reporting both high capture coverage and high automaticity were excluded. Allocation considered baseline recording practices, habit measures, and acceptance of the support features. Acceptance ratings were at least 6 out of 7 for phone reminders in PC and sharing with the researcher in SS. Table~\ref{tab:participant_baseline} reports individual characteristics and group summaries at baseline.

Participants provided informed consent and could withdraw at any time. Photos and text were shared with the research team only with participants' permission. Participants were offered USD 45 for the full study. To reduce the potential motivating effect of compensation, we emphasized that payment did not depend on capture frequency or habit formation outcomes. The study was approved by the Institutional Review Board (IRB) of our university.

\subsection{Procedure}

\begin{figure}[t]
    \centering
    \includegraphics[width=\linewidth]{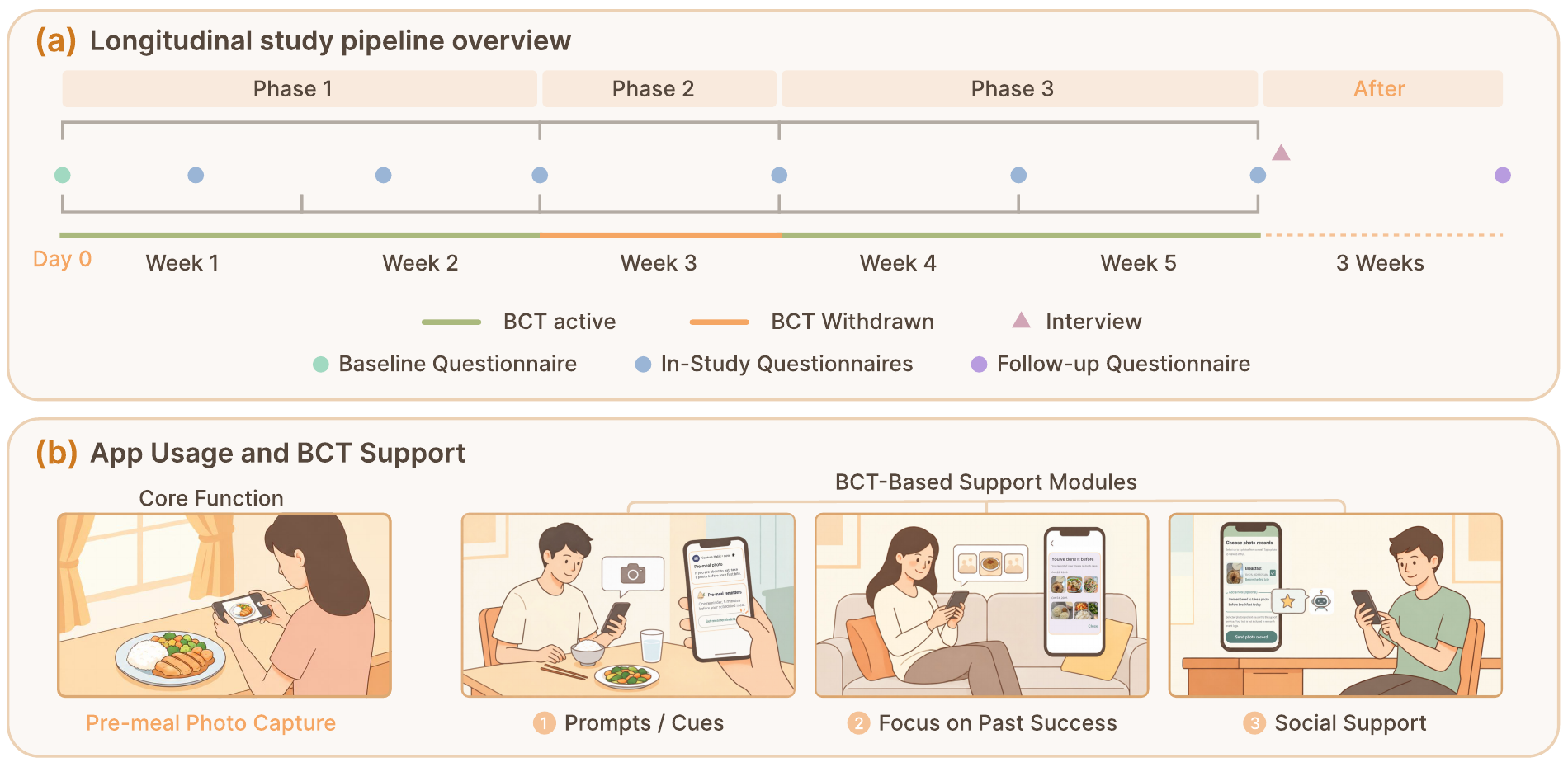}
    \caption{Longitudinal study procedure and support conditions. (a) Assigned support was available on Days~1--14, withdrawn on Days~15--21, and reinstated on Days~22--35. The timeline shows questionnaire and interview time points, followed by an unannounced follow-up three weeks after deployment. (b) All groups used the shared photo-capture functions with their assigned BCT-based support.}
    \label{fig:study_procedure}
\end{figure}

The study procedure is shown in Figure~\ref{fig:study_procedure}. During onboarding (Day~0), participants completed the baseline questionnaire, installed the app version for their assigned condition, and received training on photo capture, classification, archive use, and the assigned support module. We explained that the target behavior for habit formation was pre-meal photo capture. Participants remained in the same condition throughout the study.

The assigned support was available during Phase~1 (Days~1--14), withdrawn during Phase~2 (Days~15--21), and restored on Phase~3 (Days~22--35). The shared capture, classification, and archive functions remained available throughout. We used the initial support phase to observe changes in capture behavior and habit automaticity~\cite{stawarz2015beyond}. During withdrawal, we examined whether capture continued without additional support~\cite{lyngs2020just}, while reinstatement allowed us to examine how participants resumed using the original support.

To reduce the potential for questionnaires to act as reminders of the target behavior, we limited their frequency during deployment.
Apart from questionnaire administration, we contacted participants only in response to support requests from the Social support condition. After the five-week deployment, we conducted semi-structured interviews to discuss experiences across the three phases. A follow-up questionnaire was administered three weeks after the deployment ended.

\subsection{Data Collection}

\subsubsection{Questionnaires}

\paragraph{Pre-study questionnaire.}
Before deployment, participants completed a pre-study questionnaire. We characterized baseline habits of pre-meal photo capture using five measures that assessed the perceived naturalness of performing the behavior, recent pre-meal capture proportion, capture automaticity, spontaneous remembering, and the need for deliberate self-reminding.
Recent pre-meal capture proportion referred to the self-reported proportion of main meals photographed before the first bite during the preceding two weeks.
Capture automaticity was assessed using the four-item Self-Report Behavioural Automaticity Index (SRBAI)~\cite{gardner2012towards}. The capture proportion was reported using predefined response categories, while the remaining habit-related items used seven-point ratings.
We also measured the personal meaning and perceived value of photo recording, its perceived burden, and confidence in performing pre-meal photo capture in three recurring contexts. These measures provided a baseline for subsequent repeated assessments. We collected contextual information, including mealtime regularity, to help interpret variation in capture behavior and support experiences.

\paragraph{Questionnaires during deployment.}

Participants completed deployment questionnaires on Days~4, 9, 14, 21, 28, and 35. These questionnaires assessed capture automaticity (SRBAI), spontaneous remembering, the need for deliberate self-reminding, the personal meaning of photo recording, and perceived burden. At selected time points, additional questions assessed the perceived value of photo recording, integration into everyday meal routines, confidence in challenging meal contexts and in performing the behavior independently, and recording motivations.
During Phase~1 and Phase~3, condition-specific questions assessed use of and experiences with the assigned support. For Prompts/cues, these questions covered reminder timing, helpfulness in remembering to capture pre-meal photos, perceived intrusiveness, and whether reminders were increasingly ignored. 
For Social support, they covered feeling seen and supported, experiences of waiting for responses, the burden of sharing, and willingness to engage. For Focus on past success, they covered the meaning attributed to successful capture records, confidence in repeating the behavior, and changes over time in the perceived helpfulness of reviewing these records. Phase~2 questions addressed awareness of support withdrawal, expectations, and coping responses during withdrawal, while questions for Phase~3 assessed experiences after reinstatement and changes in support needs. Additional questions examined the active effort required to sustain capture, sources of help participants considered effective, and preferences for future recording and support. To cross-check meal counts, participants also reported skipped main meals and photos taken with other tools.

\paragraph{Follow-up questionnaire.}
After deployment ended, we administered a follow-up questionnaire to examine the continuation of pre-meal photo capture. Participants were not informed about the follow-up in advance to reduce the likelihood that anticipation of the assessment would influence their recording behavior. After reviewing their records from the preceding seven days, participants reported the total number of main meals and the number of pre-meal photos. They also reported any changes in recording behavior after the field study.

\subsubsection{Post-study interviews}
At study completion, we conducted semi-structured interviews to understand how participants’ capture behavior and subjective experiences changed across these phases. Interviews used shared reflection, and questions were informed by their questionnaire responses and records. Interviews explored recording goals and perceived habit changes, the benefits and limitations of initial support, coping after withdrawal, and changes in support use and needs following reinstatement. We also discussed how time pressure, eating with others, changing mealtimes, and recording tools influenced capture, and which aspects of support participants wished to retain, modify, or remove.

\paragraph{Application usage logs}
The app automatically logged timestamped events for photo capture, saving, deletion, and annotation changes. Photo records included participants' annotations of meal type and pre-meal status. Support logs for the Prompts/cues condition recorded reminder settings, delivery, taps, and dismissals. For the Social support condition, the logs recorded request submissions and the availability, display, and expansion times of replies. For the Focus on past success condition, they recorded review material selection, review openings and duration. Additional study-related social support exchanges via WeChat were documented separately.

\subsection{Data Preprocessing and Analysis}



\subsubsection{Quantitative Analysis}

All 54 participants contributed data throughout the initial support and withdrawal phases. Four participants chose not to continue into Phase~3, reporting that they felt they had achieved their habit-formation goals. Their earlier observations were retained. Logs and questionnaires were aligned with each participant's study timeline. Actual main-meal opportunities (\(E\)) were reconstructed from a baseline of three meals per day, adjusted using reported deviations. \(T\) was the number of meals with a valid capture at any time, and \(P\) was the number confirmed as captured before the first bite. Both counts were deduplicated at the meal level. The primary behavioral outcome was \(P/E\), with \(T/E\), \((T-P)/E\), and daily \(P\) volume as supplementary outcomes. Weekly and phase-level proportions were calculated for each participant and then averaged, giving each participant equal weight.

Capture automaticity was calculated as the mean of the four SRBAI items. Spontaneous remembering and the need for deliberate self-reminding retained their original 1--7 scoring directions. Contextual confidence was calculated as the mean of three meal-context ratings. We constructed an equally weighted baseline habit index from perceived naturalness, recent pre-meal capture proportion, SRBAI, spontaneous remembering, and low effort after rescaling each dimension to 0--100. We rank-ordered the index in the original 54 participants to define fixed, equal-sized low, medium, and high strata (\(n=18\) each), which were retained for all stratified analyses. 
Separately, we standardized the index against the original sample and used it as a continuous variable for adjustment and moderation. Predictions at \(z=-1\) and \(z=+1\) refer to this continuous score.

Pre-meal capture adherence was analyzed using binomial-logit mixed models with \(P\) as successes, \(E\) as opportunities, participant-level random intercepts, and participant-clustered robust inference. The models included support condition, categorical week or phase, their interaction, and the continuous baseline index. Moderation models added the full condition \(\times\) time \(\times\) baseline-index interaction. The initial support phase was divided into D1--4, D5--7, D8--9, and D10--14. Full-study models used W1--W5. We report marginal predicted probabilities and direct contrasts in levels and changes, including estimates at \(z=\pm1\). Three-phase descriptive comparisons used the same 50 participants and paired participant-level bootstrap intervals. Daily \(P\) volume was analyzed using Poisson GEE with condition, phase, their interaction, and the continuous baseline index, without an exposure offset. Segmented binomial-logit GEE modeled condition-specific within-phase daily trends, adjusting for the continuous baseline index. Both analyses used participant-robust variance estimates.

The three core subjective outcomes were analyzed across seven assessments using linear mixed-effects models that included support condition, categorical assessment, their interaction, and participant-level random intercepts and linear time slopes. Baseline-to-D35 contrasts used data from the 50 participants who continued into reinstatement. Early changes by fixed baseline stratum were analyzed after pooling support conditions, using participant-clustered linear models across baseline, D4, D9, and D14. An additional random-intercept model tested the condition \(\times\) assessment \(\times\) continuous baseline-index interaction for the need for deliberate self-reminding at D14, D21, and D35. The model adjusted for the baseline score of this outcome and the baseline-by-assessment interaction. The three-way interaction was evaluated using a likelihood-ratio test. Baseline-to-D14 change in contextual confidence was analyzed using a paired \(t\)-test.

At the three-week follow-up, \(P/E\) was compared across conditions for all 54 respondents using fractional-logit regression adjusted for the continuous baseline index, with robust covariance estimation and standardized marginal predictions. The three pairwise condition comparisons used Holm adjustment. Changes from Week~5 to follow-up for the 50 paired participants were estimated using paired participant-level bootstrap intervals, with verified deployment logs and record-assisted seven-day reports collected at follow-up treated as distinct data sources.

\subsubsection{Qualitative Analysis}
We conducted semi-structured interviews with 22 participants, including 2 who ended participation after Phase~2 and 20 who completed all three phases. Other participants provided written feedback on similar topics.
We conducted an inductive thematic analysis of the interview transcripts and open-ended questionnaire responses from the supplementary and follow-up questionnaires. The first author independently coded the data and discussed the analysis with the research team in relation to our RQs.

%% file: section/6_results.tex
\section{Results and Findings}


\subsection{Capture trajectories and changes in perceived habit}
\label{sec:trajetory_and_changes}

\subsubsection{Five-week capture trajectories}
\label{sec:capture_trajetory}

\begin{figure}[h]
    \centering
    \includegraphics[width=\linewidth]{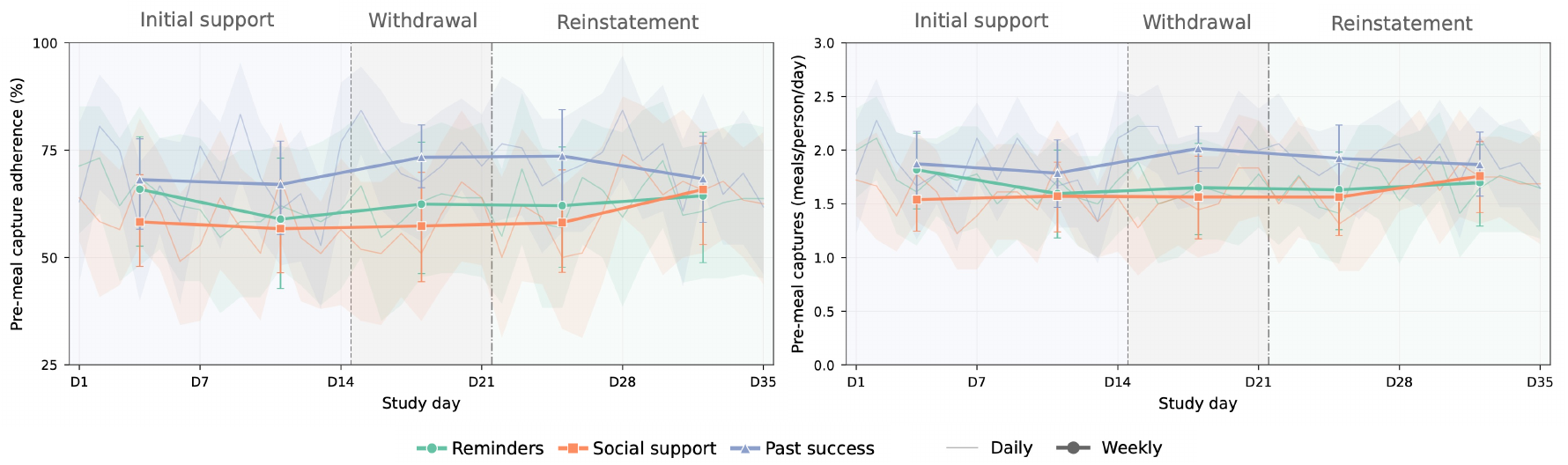}
    \caption{Daily trajectories and weekly summaries of pre-meal capture by support condition. The panels show adherence ($P/E$, left) and pre-meal capture volume ($P$, meals per participant per day, right).}
    \label{fig:premeal_adherence}
\end{figure}

Across all participants, records covered 4,945 main meals. 70.48\% (3,485/4,945) had capture records, and 63.92\% (3,161/4,945) were confirmed pre-meal captures. Figure~\ref{fig:premeal_adherence} presents daily trajectories and weekly summaries of pre-meal capture adherence and pre-meal capture volume for each BCT support condition.

Pre-meal capture was sustained across all study phases. Mean participant-level adherence was 63.8\% during initial support, 66.3\% during withdrawal and 65.1\% during reinstatement. 
Corresponding daily pre-meal capture volumes averaged 1.74, 1.79, and 1.74 meals per participant. Overall adherence did not change significantly from initial support to withdrawal or from withdrawal to reinstatement. 

The weekly trajectories showed different descriptive patterns across the three conditions. Adherence in the Prompts/cues condition fell in Week~2 before recovering, adherence in the Social support condition dipped in Week~4 before rising, and adherence in the Focus on past success condition peaked during withdrawal before declining. The three conditions showed modest, nonparallel fluctuations, with no common withdrawal drop or reinstatement rebound. By Week~5, adherence was similar across conditions, at 64.4\%, 65.8\%, and 68.3\%.

Although condition-by-week interaction in pre-meal capture adherence was not significant, the Focus on past success condition showed higher pre-meal capture adherence than the Social support condition ($\Delta=15.09\%$, 95\% CI [0.57, 29.60], $p=.042$) in Phase~2. The volume of daily pre-meal capture was also higher in the Focus on past success condition ($\Delta=0.453$, 95\% CI [0.013, 0.892], $p=.044$). These Phase~2 differences suggest a possible advantage of Focus on past success in promoting and maintaining the behavior during habit formation.

The mean proportion of meals that were captured but not confirmed as pre-meal decreased from 7.89\% in Week~2 to 3.63\% in Week~5. This decrease indicates a gradual shift from remembering after eating had begun to completing capture before the meal.

\subsubsection{Stronger automaticity and less need for deliberate self-reminding}

\begin{figure}[h]
    \centering
    \includegraphics[width=\linewidth]{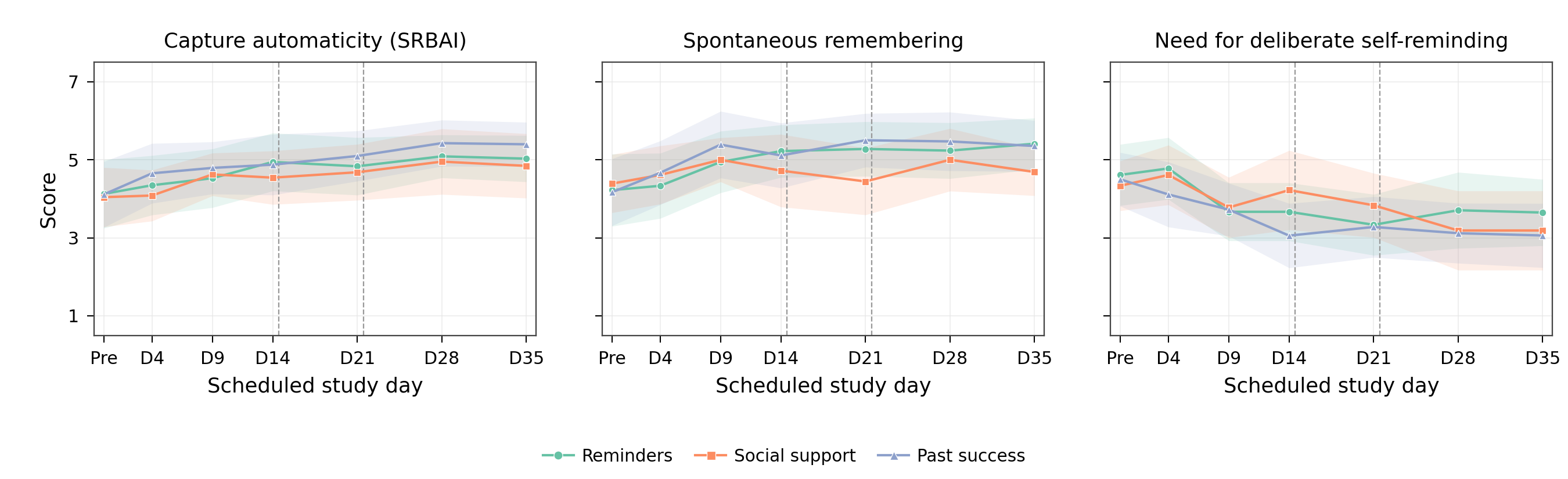}
    \caption{Trajectories of capture automaticity (SRBAI), spontaneous remembering, and the need for deliberate self-reminding by support condition.}
    \label{fig:questionnaire_core}
\end{figure}

Participants who completed all three phases reported increased automaticity and spontaneous remembering and a reduced need for deliberate self-reminding, as shown in Figure~\ref{fig:questionnaire_core}. Capture automaticity increased from 4.17 at baseline to 5.10 at D35 (95\% CI [0.396, 1.464], $p<.001$). Spontaneous remembering increased from 4.32 to 5.16 (95\% CI [0.217, 1.463], $p=.008$), while the need for deliberate self-reminding decreased from 4.50 to 3.30 (95\% CI [$-1.886$, $-0.514$], $p<.001$). 
Much of the improvement was evident by the end of Phase~1, with no significant further changes during withdrawal or reinstatement.

Condition-by-assessment interactions were nonsignificant for automaticity, spontaneous remembering, and the need for deliberate self-reminding.
Although improvements remained evident in later phases, spontaneous remembering in the Social support condition declined toward the end of initial support and during withdrawal, rebounded in the first week of reinstatement, and declined again at D35. This descriptive pattern may point to a limited role for this form of social support in sustaining self-initiation over time.

\subsubsection{Initial habit levels suggested different roles for support}

\begin{figure}[h]
    \centering
    \includegraphics[width=\linewidth]{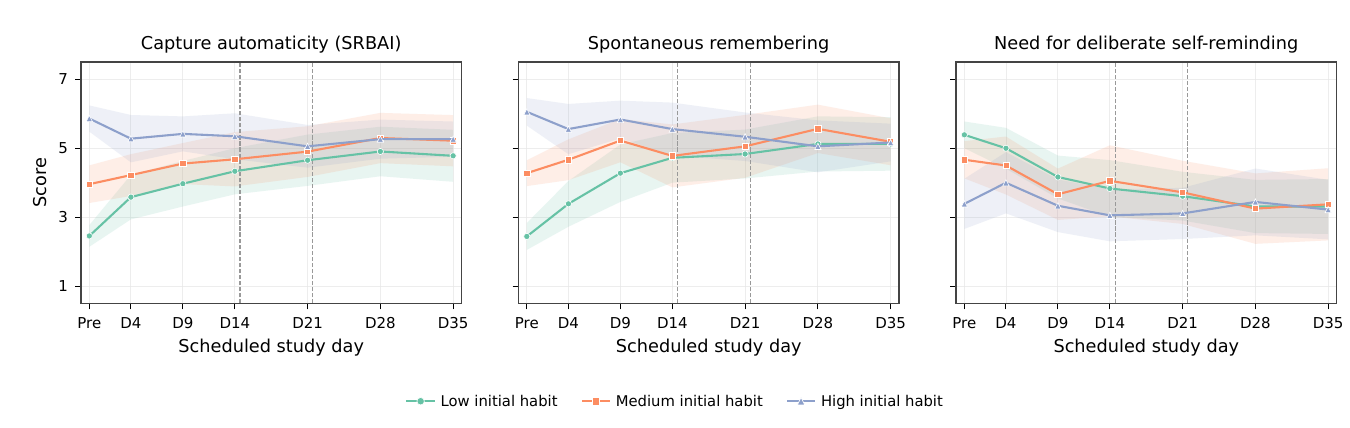}
    \caption{Trajectories of three core subjective habit measures by initial habit stratum, pooled across support conditions. Low, medium, and high strata were defined using the five-dimensional baseline habit index and remained fixed throughout the study.}
    \label{fig:initial_habit_subjective}
\end{figure}

Pooling data across the three support conditions, the low initial-habit stratum showed larger early improvements than the high stratum. From baseline to D9, the between-stratum differences in change were 1.958 points for automaticity (95\% CI [1.197, 2.720], $p<.001$) and 2.056 points for spontaneous remembering (95\% CI [1.038, 3.073], $p<.001$), narrowing the initial gaps in perceived habit, as shown in Figure~\ref{fig:initial_habit_subjective}. Descriptive trajectories for each support condition within the initial-habit strata are provided in Appendix~\ref{app:longitudinal_study}. Across D14, D21, and D35, the need for deliberate self-reminding also showed a support condition $\times$ phase $\times$ initial habit interaction ($\chi^{2}(4)=9.639$, $p=.047$), indicating that its association with initial habit changed differently across support conditions.

\begin{figure}[h]
    \centering
    \includegraphics[width=\linewidth]{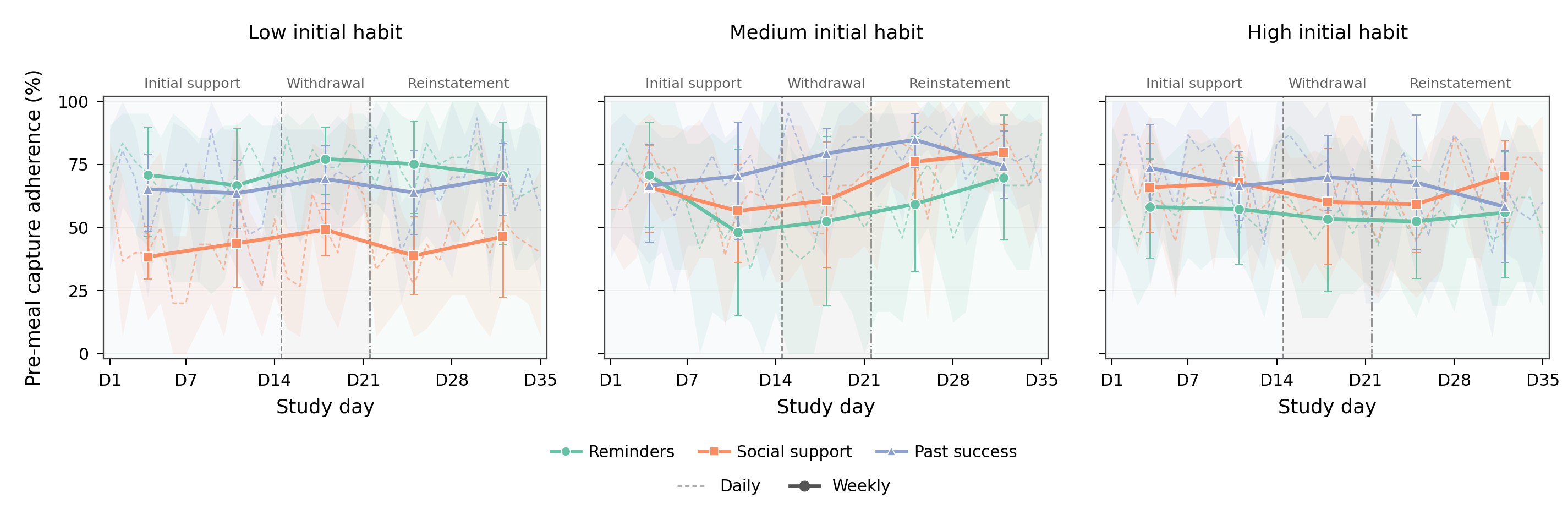}
    \caption{Daily and weekly mean pre-meal capture adherence ($P/E$) by support condition within each initial habit stratum. The panels show the low, medium, and high strata, with membership fixed at baseline.}
    \label{fig:composite_daily_weekly_adherence}
\end{figure}

\begin{figure}[h]
    \centering
    \includegraphics[width=\linewidth]{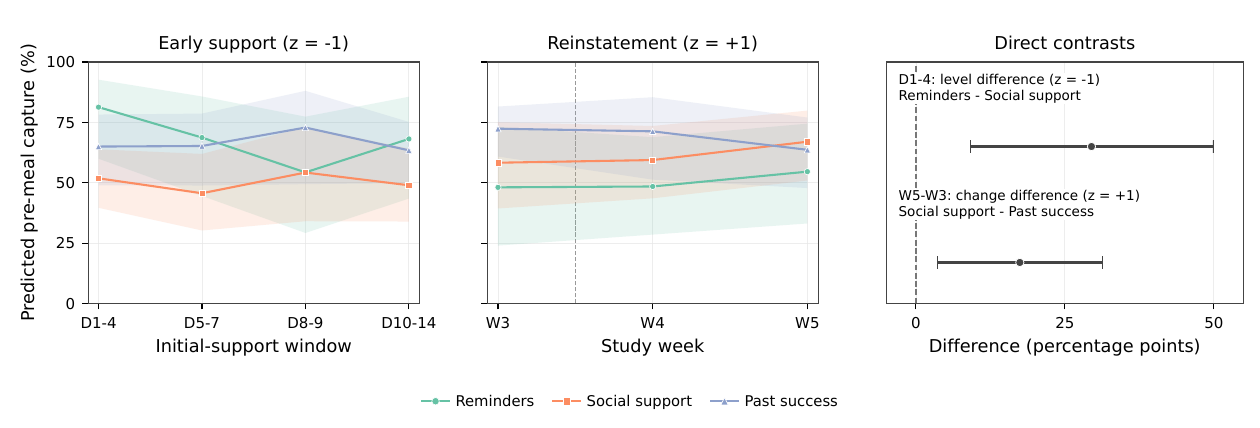}
    \caption{Model-estimated pre-meal capture adherence at initial habit scores one SD below ($z=-1$, left) and above ($z=+1$, middle) the mean. The right panel shows the corresponding between-group contrasts in adherence levels and changes.}
    \label{fig:early_late_conditional_adherence_with_contrasts}
\end{figure}

Although the overall five-week three-way interaction for adherence was nonsignificant, conditional comparisons suggested different roles for support. In the low initial-habit stratum, the Prompts/cues condition showed the highest observed weekly adherence across all five weeks (66.7\%--77.1\%, Figure~\ref{fig:composite_daily_weekly_adherence}), with the clearest separation from the Social support condition. At an initial habit score one $SD$ below the mean, estimated pre-meal capture adherence during D1--4 was higher in the Prompts/cues condition than in the Social support condition ($\Delta=29.46\%$, 95\% CI [9.09, 49.84], $p=.005$), as shown in Figure~\ref{fig:early_late_conditional_adherence_with_contrasts}.
SS16, in the low initial-habit stratum, noted that using social support required \textit{"first remembering to open the app"}. This account indicates that direct reminders could address an initiation difficulty faced by participants in the low initial-habit stratum.

Descriptively, the Focus on past success condition showed the highest observed weekly adherence in the medium stratum during Weeks~2--4 (70.3\%, 79.3\%, and 84.7\%, Figure~\ref{fig:composite_daily_weekly_adherence}). PS4, in this stratum, described previous captures as evidence that they could \textit{"keep doing it"}. Reviewing completed captures could thus make continued effort feel achievable by grounding confidence in participants' own experience.

At an initial habit score one $SD$ above the mean, adherence followed opposing trends from withdrawal (Week~3) to the final week of reinstatement (Week~5). Estimated adherence increased from 58.29\% to 66.99\% in the Social support condition and decreased from 72.39\% to 63.65\% in the Focus on past success condition. The between-condition difference in change was 17.44\% (95\% CI [3.55, 31.32], $p=.014$), as shown in Figure~\ref{fig:early_late_conditional_adherence_with_contrasts}.
These opposing trends narrowed the earlier advantage of Focus on past success. SS14, in this stratum, said that after reinstatement, sharing made them feel that \textit{"someone was seeing my records"}, even without an immediate reply. This account suggests that social support could sustain motivation for continued capture even when the behavior was already familiar.

\subsection{Continued effort and support needs as capture became familiar}

\subsubsection{Spontaneous remembering coexisted with deliberate strategies}

Continued capture during withdrawal reflected both growing familiarity and active efforts to remember. Adherence was broadly sustained, as shown in Section~\ref{sec:capture_trajetory}. In the Social support condition, estimated adherence increased by 16.80\% from the first to the last day of withdrawal (95\% CI [4.12, 29.48], $p=.009$).

In retrospective comparisons, the proportion of participants reporting spontaneous remembering increased from 26/54 for D10--14 in initial support to 43/54 for withdrawal phase. PC15 described capture \textit{"moving from a task to part of my everyday life"}, suggesting that capture was becoming integrated into meal routines. Such integration provided a basis for continuation when the assigned support was unavailable.

Deliberate strategies remained part of this increasingly familiar practice. During withdrawal, 32/54 participants reported consciously reminding themselves, and 16/54 relied on familiar contextual cues. 23 participants reported both spontaneous remembering and deliberate self-reminding. This overlap suggests that the same participant could remember readily at some meals while using deliberate strategies to sustain capture when familiarity alone was insufficient.
Without study prompts, PC2 needed alternatives such as \textit{"setting alarms"} or \textit{"writing it in my schedule"}, yet could still turn off an alarm and forget to take a photo. Support withdrawal could therefore shift some of the work of sustaining capture to participants themselves.

\subsubsection{Support availability was valued despite infrequent active use}

Even as active interactions declined, participants continued to value access to support. From Week~1 to Week~2, logged interactions decreased across all three features, from 36 to 20 in the Prompts/cues module, from 107 to 39 requests in the Social support module, and from 40 to 19 reviews in the Focus on past success module. In Weeks~4 and~5, after reinstatement, logs recorded 33 and 28 in the Prompts/cues module, 41 and 32 in-app requests in the Social support module, and 15 and 25 reviews in the Focus on past success module.
At the end of the five-week study, 26 participants felt reassured knowing that support remained available even when they used it infrequently (social support: $n=11$; reminders: $n=8$; past success: $n=7$).
Regaining needed help and feeling able to record in their own way with little need for support were each reported by 18 participants. Eight participants reported both reassurance and a personal way of recording. This overlap suggests that developing a personal recording routine can coexist with valuing access to BCT-based support when needed.

Participants' accounts illustrated different meanings of this reassurance. PC7, who could usually remember to capture without prompts, described the reminder as a \textit{"backup"} and said that \textit{"seeing it made me feel reassured"}. SS1 emphasized that this support served as a channel for expressing emotions. SS7 similarly explained that although capture continued during withdrawal, the process felt \textit{"a little lonely"}. When sharing support was restored, it made the experience feel more \textit{"secure and warmer"}. The opportunity to share continued to shape the recording experience even when only selected meals were shared.

These patterns suggest that support could become supplementary as capture became familiar. Prompts/cues offered an additional safeguard against forgetting, while social support preserved opportunities for expression and companionship. The continued availability of BCT-based support could provide reassurance beyond occasions of active use during habit formation.

\subsection{Usefulness of support depended on fit with everyday practice}

\subsubsection{Remembering still had to be translated into feasible action}
Alongside gains in perceived automaticity, participants reported lower confidence in pre-meal capture when rushed, when mealtimes changed, or when eating with others. Mean confidence across these situations decreased by 0.833 points from baseline to D14 (95\% CI [$-1.253$, $-0.413$], $p<.001$). PC6 prepared for capture when a reminder arrived while cooking, retaining the notification to open the app once the meal was ready. This made \textit{"photographing before eating smoother"}. 
For PC7, changed mealtimes without corresponding settings updates could leave reminders too early, and they might forget to capture by mealtime. PS9 described how rushed meals or eating while on duty disrupted the usual capture routine.
These accounts suggest that easier remembering and an emerging habit did not guarantee completion. Capture still depended on timing and opportunities to act, which competing demands on attention could disrupt. Prompts/cues could redirect attention in the moment but could not ensure an opportunity to act.
Such disruptions coexisted with overall gains in perceived automaticity, suggesting that participants could still experience the behavior as increasingly habitual even when capture was interrupted in special situations.

\subsubsection{Value of social support depended on expectations}
The value of social support depended on who they shared with and what they expected from the exchange. Participants valued personally relevant responses, freedom from judgment, and control over the interaction.
SS1 expected varied replies but found responses predictable, which sometimes reduced their willingness to share. He felt that family could respond more appropriately because they knew him well. SS12 similarly felt cared for when family members asked about their meals and nutrition. 
Such questions could also prompt SS12 to take and share a meal photo, connecting recording with everyday family communication.
Moreover, more specific or immediate responses were not necessarily preferred. SS2 felt that generic replies offered limited encouragement, but appreciated not having to \textit{"worry about being judged"}. SS7 valued asynchronous replies and the freedom to decide whether and when to share, without pressure to respond. Having initially worried about criticism of their meals or pressure to record, SS7 felt that real-time conversation \textit{"might instead become a burden"}. These accounts suggest that making sharing more personally meaningful and keeping interaction manageable were both relevant to how social support fitted participants' recording practices.

\begin{figure}[b]
    \centering
    \includegraphics[width=\linewidth]{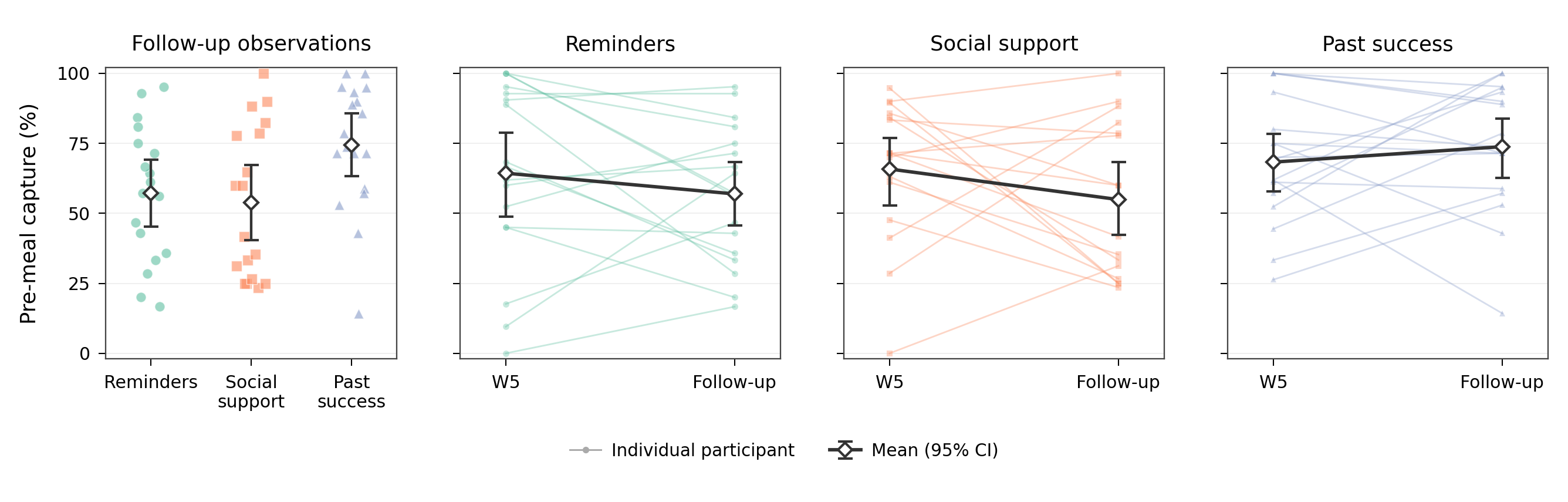}
    \caption{Observed pre-meal capture adherence at follow-up and changes from Week 5, by support condition. The left panel shows all follow-up observations. The remaining panels show paired observations for participants with data at both time points. Follow-up proportions were based on record-assisted reports for the preceding seven days.
    }
    \label{fig:followup_distribution_and_paired_change}
\end{figure}

\subsubsection{Revisiting records served purposes beyond reinforcing confidence}
Accounts from participants in the Focus on past success condition showed that accumulated records could acquire additional value and provide further reasons to continue capturing meals. As the need to reinforce confidence in persistence became less central, the appeal of reviewing could shift toward appreciating the experiences preserved in the photos and using the records.
PS4 drew confidence from previous captures, reasoning that having recorded meals before meant they could continue. PS5 valued the memories and feelings evoked by meal photos. One selected meal photo brought back a relaxed and enjoyable moment, while browsing the growing photo diary brought together \textit{"fragments of my everyday life"}. PS5 felt that a single fixed image offered limited support compared with a collection they could browse. As the diary grew, PS5 described moving from completing a recording task toward wanting to preserve everyday moments for later review.
PS18 described a change in emphasis over time. Looking back initially helped reinforce confidence in continuing, but during reinstatement, \textit{"that confidence was already there"}, and reviewing became more about looking back on what they had eaten and how healthy their daily diet was.

\subsection{Continued recording took different forms after deployment}
\label{sec:after_study_recording}

At follow-up, participants reviewed their records and reported main meals and pre-meal captures for the preceding seven days. Across all respondents, adherence after adjustment for initial habit was highest in the Focus on past success condition. It exceeded adherence in the Prompts/cues condition by 17.72\% (95\% CI [3.38, 32.06], $p_{Holm}=.031$) and adherence in the Social support condition by 20.54\% (95\% CI [5.05, 36.04], $p_{Holm}=.028$). Among these completers, overall mean adherence changed from 66.2\% in Week~5 to 62.0\% at follow-up.

\begin{figure}[h]
    \centering
    \includegraphics[width=.85\linewidth]{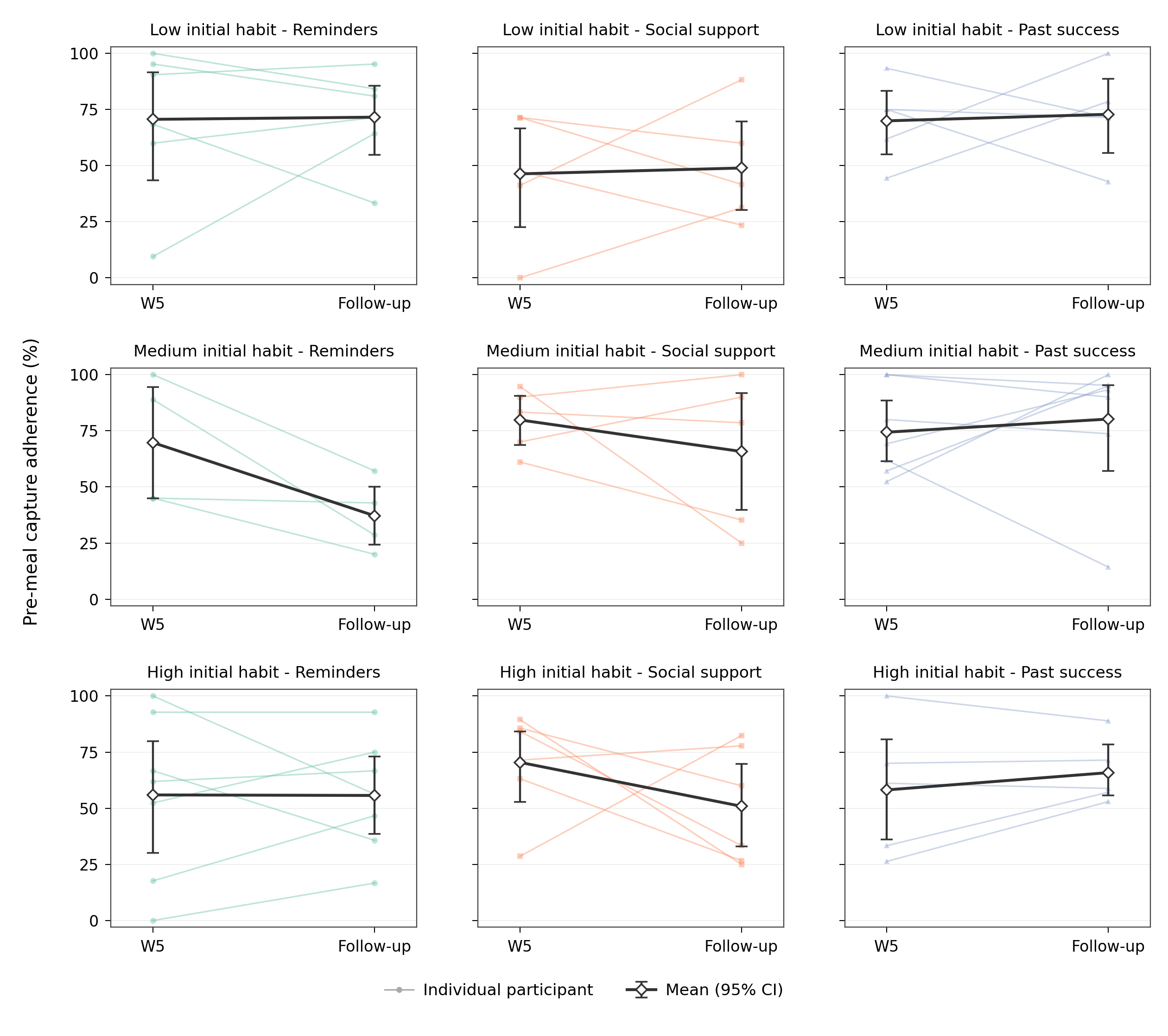}
    \caption{Paired changes in pre-meal capture adherence from Week~5 to follow-up by support condition and fixed initial-habit stratum. Comparisons are shown by support condition and habit stratum, with stratum membership retained from baseline. Each time point represents a seven-day window, with follow-up proportions based on record-assisted self-reports.
    }
    \label{fig:composite_stratified_followup_pairs}
\end{figure}

The Focus on past success condition showed the most favorable continuation pattern. Mean adherence increased by 5.53\%, whereas it decreased by 7.42\% in the Prompts/cues condition and by 10.92\% in the Social support condition. Figure~\ref{fig:composite_stratified_followup_pairs} further shows how these patterns varied across initial habit strata.
Within the Focus on past success condition, mean adherence increased by 2.9\%, 5.8\%, and 7.7\% in the low, medium, and high strata. The largest declines occurred in the medium stratum of the Prompts/cues condition ($-32.6\%$), the medium and high strata of the Social support condition ($-14.1\%$ and $-19.6\%$). In the low stratum, the Prompts/cues and Focus on past success conditions retained similar follow-up levels ($71.6\%$ and $72.9\%$). These exploratory trajectories suggest that the favorable performance of Prompts/cues among participants with low initial habit continued after assigned prompting ended, while the relative advantage of Focus on past success was more apparent in the medium and high strata.

The personal value participants had found in accumulated records offers one possible explanation for continued capture. At follow-up, PS13 changed recording tools but retained \textit{"the desire to photograph my main meals"}. PS18 reported little change in capture frequency despite reviewing less often. These accounts suggest that reasons to record could persist beyond the original feature, even when reviewing became less frequent. SS9, however, missed feeling that \textit{"someone was keeping up the practice with me"} and attributed reduced capture to forgetting and feeling less motivated when recording alone. Continuing the behavior could therefore involve retaining personally meaningful reasons to record, while motivation rooted in shared participation could weaken when that arrangement ended.
Continuation also involved choices about what to record and how to use the photos. SS1 selected memorable meals and would \textit{"not record everything anymore"}. PS17 reported photographing almost every meal with the native camera \textit{"without any reminders"} and sending the photos to their partner. Recording could thus remain personally meaningful as its coverage changed, with familiar tools and personal relationships providing ways to carry it forward.

%% file: section/7_discussion.tex
\section{Discussion}


Across three BCT-based supports differing in intervention timing, required user initiation, and resource use, pre-meal capture was broadly maintained across support, withdrawal, and reinstatement. Differences emerged by initial habit level and study phase, but no condition showed a consistent advantage throughout the study \textbf{(RQ1)}. Early improvements in perceived automaticity largely persisted, but support effects did not follow a uniform progression from external assistance to independent capture. As capture became more familiar, participants continued to use deliberate strategies and value the availability of support even as active feature use declined \textbf{(RQ2)}. Participants' experiences indicated that support was useful when it helped them remember at feasible moments, met their interaction expectations, and allowed accumulated records to serve changing purposes \textbf{(RQ3)}.
We therefore argue that longitudinal BCT design involves two coupled problems: matching support to the need that currently constrains action, and reconfiguring where the work of sustaining capture is located across users, system cues, personal records, and support relationships.

\subsection{Configuring BCT support around changing needs in pre-meal capture}
Pre-meal photo capture requires little effort but is sensitive to timing and vulnerable to forgetting and irregular meal schedules~\cite{cordeiro2015rethinking,lee2024tailored}. Our findings highlight the difficulty of turning an intention to record into action before the first bite.

Participants already had recording intentions and generally knew how to take a photo. Among those with lower initial habit levels, the difficulty appeared to be shifting attention to capture within the narrow pre-meal window, while the ease of acting still varied across meal contexts. This pattern suggested a role for reminders when capture was not yet reliably brought to mind.
Noticing a prompt and completing capture remained distinct. A cue could prepare users for later action when connected to the upcoming meal but lose relevance if delivered too early. The Fogg Behavior Model helps explain this distinction~\cite{fogg2009behavior}. Behavior occurs when motivation, ability, and a prompt converge at the relevant moment. In this setting, participants generally knew how to take a photo, but their situational ability varied with food preparation, meal pickup, conversation, and the onset of eating. Prompt design should therefore connect capture to a meal transition when action is feasible and minimize the steps from cue to capture.

The three BCT implementations addressed different constraints in this process. Prompts/cues supported timely remembering and reduced the steps to capture. Focus on past success turned prior records into self-generated evidence that could support confidence and continued action. Social support offered recognition, reassurance, or connection. Support fit therefore depended on both the kind of help offered and whether that help was accessible when needed~\cite{klasnja2011evaluate,hekler2013mind}. A feature that strengthened confidence or motivation could still offer limited help when the immediate barrier was remembering to begin.


Early gains in perceived automaticity and spontaneous remembering were largely sustained as active feature use declined, resembling trajectories reported for nudging interventions~\cite{he2026does}. Accumulated photo records also came to support meal recall, reflection on eating, and the preservation of personally meaningful experiences. These records became part of participants' evolving goals, tools, and tracking practices, highlighting the distinction between engagement with a feature and engagement with the target behavior~\cite{alberts2026designing}. Declining feature use may reflect greater familiarity, reduced need, a change in purpose, or a shift toward occasional use. Interaction frequency should therefore be interpreted in relation to the need that a feature currently serves.
These findings suggest a configurable support pathway. Prompts/cues can support timely remembering when initiation is the main barrier. As users accumulate experience, Focus on past success can surface self-generated evidence when confidence or reasons to continue become more important. Prompting and social support can remain available when changes in routines or circumstances make timely remembering, reassurance, or connection salient again. This pathway is reversible and does not assume a fixed sequence of stages.

The main design challenge is deciding which system elements should remain stable as support changes. Meal-related cues, the camera entry point, and the sequence from cue to capture can remain predictable to reinforce associations between context and action~\cite{stawarz2015beyond}. Other support elements can remain low burden and under user control~\cite{consolvo2009theory}. Systems can adjust reminder timing and frequency, the records surfaced for reflection, and the availability or intensity of social interaction. Intervention rotation presents a trade-off because variation may renew behavioral impact but also increase attrition~\cite{kovacs2018rotating}. These adaptations should preserve the established sequence from meal cue to capture. We describe this principle as \textbf{preserving a predictable capture pathway while adapting support content and intensity}.

\subsection{Sustaining pre-meal capture through changing support arrangements}


Beyond the needs addressed by each support, a related question is how the work of sustaining capture was distributed and reorganized. Pre-meal capture may appear to be an individual habit, but maintaining it partly depended on a sociotechnical arrangement of cues, interfaces, personal records, and support relationships. Prompts/cues shifted part of the remembering and timing work to the system, while users configured and acted on the cues. Focus on past success located part of the confidence-building and interpretive work in accumulated records, while users initiated review and connected past experiences to current needs. Social support distributed relational work across the participant, the interface, and a researcher or chatbot, while users initiated sharing and managed response expectations. These implementations represented different emphases within a shared practice, not mutually exclusive categories of system, self, or human mediation. A BCT implementation can reduce one form of work while relocating another, so its role depends on who acts, when, and with what effort.

Withdrawal and reinstatement made this allocation of work visible. During withdrawal, broadly maintained capture coexisted with more spontaneous remembering and continued deliberate strategies. Some participants supplemented the removed support with personal alarms, schedules, and familiar meal cues, so similar behavioral outcomes could rest on different combinations of routinized action, active self-reminding, and external resources. Withdrawal did not simply test whether capture persisted; it reconfigured who or what carried the work of sustaining it. Reinstatement then reintroduced support into an arrangement that participants had already modified, sometimes as direct assistance and sometimes as a backup resource. We therefore interpret habit support at two levels: helping capture become more self-initiated, and enabling the broader arrangement that sustains repeated action to be reorganized. Support reduction should be designed as a reversible reconfiguration, allowing users to connect their own strategies and resources while retaining a clear route back to assistance when routines change.

Support availability emerged as a design dimension distinct from proactive intervention frequency. Participants could interact less with reminders, review, or sharing while continuing to value reassurance, an avenue for expression, or the possibility of being supported. Low use and low value were therefore not equivalent in our data. Systems can configure these properties separately: users can reduce or pause reminders while retaining a visible reactivation entry; share less often while preserving a communication channel with clear, low-pressure response expectations; or review less frequently while keeping access to accumulated records. Low interaction should prompt reassessment of the role a support currently plays, not its automatic removal. We describe this principle as \textbf{configuring proactive intervention separately from support availability}. It allows support to recede from everyday attention while remaining available within the arrangement through which users sustain capture.

Existing records, familiar tools, and chosen relationships can carry support beyond a particular feature or application. Participants used meal photos for memory and reflection, shifted capture to native camera apps, and continued sharing through personal relationships after deployment. These patterns add a target-behavior perspective to lived-informatics accounts of changing goals and tools~\cite{epstein2015lived} and to accounts of socially sustained self-tracking, in which technologies and social resources can be combined, substituted, and resumed over time~\cite{lu2021model}. Systems can surface photos users have already produced, keep histories portable and easy to revisit, and connect sharing to people and channels selected by the user. These choices reduce the additional organization and interaction required to obtain support. Continued recording through another tool or selective capture is not equivalent to completing pre-meal capture at every main meal. Longitudinal evaluation should therefore consider both the personal continuity of the practice and adherence to the specified target behavior. We summarize this principle as \textbf{building on existing records and chosen relationships to reduce the work of obtaining support}. This principle treats BCT support as a configurable sociotechnical arrangement and shifts the unit of longitudinal evaluation from a fixed feature to the evolving arrangement through which the target practice is sustained. Evaluation should trace both behavioral persistence and how support work is taken up, substituted, and reactivated across people and artifacts~\cite{klasnja2011evaluate}.

\subsection{Limitations and Future Work}
This study examined one implementation of each of three BCTs with 54 participants over five weeks and a three-week follow-up. The findings capture an early period of changing support use and perceived habit development. Maintenance over several months remains an open question.
Questionnaires, awareness of app logging, study contact, and compensation may also have modestly increased attention to capture despite efforts to limit these effects. Future work can extend deployment, recruit broader populations, and test BCT combinations, rotation, or adaptive sequencing as users' routines and support needs change.

%% file: section/8_conclusion.tex
\section{Conclusion}

We investigated three participant-informed BCT supports, including Prompts/cues, Social support, and Focus on past success, for supporting habit formation in pre-meal photo capture across support, withdrawal, reinstatement, and follow-up. Pre-meal capture was broadly maintained and perceived automaticity strengthened. As capture became more familiar, participants combined spontaneous remembering with deliberate strategies, interacted less with support while continuing to value its availability, and used accumulated records for confidence, reflection, and personal meaning. Withdrawal and reinstatement further showed that similar behavioral outcomes can be sustained through different arrangements of user effort, system cues, personal records, tools, and support relationships. We therefore frame longitudinal BCT design around two linked concerns: matching support to the need that currently constrains action, and allowing the work of sustaining behavior to be reconfigured over time. These findings position BCT support as a configurable sociotechnical arrangement. Future BCT-based systems should preserve a predictable capture pathway, adapt support content and availability, and build on users' existing records and relationships.

%% file: section/appendix.tex
\newpage

\section{Literature Search, Screening, and BCT Coding}
\label{app:literature_screening}

\subsection{Search Strategy}

We searched the ACM Digital Library for health behavior change articles published in CHI, DIS, UIST, IMWUT, and CSCW between January 1, 2019 and June 30, 2026. Search terms were adapted from \citet{wang2019health} and covered general behavior change terminology and specific health behaviors. The complete query is reproduced below.

\begin{tcolorbox}[
    width=\linewidth,
    colback=black!6,
    colframe=black!50,
    boxrule=0.4pt,
    arc=0pt,
    boxsep=0pt,
    left=6pt,
    right=6pt,
    top=5pt,
    bottom=5pt,
    before skip=6pt,
    after skip=6pt,
    fontupper=\footnotesize
]
\raggedright
[Keywords: behavior change] OR
[Keywords: behavioral change] OR
[Keywords: behaviour change] OR
[Keywords: behavioural change] OR
[Keywords: persuasive technology] OR
[Keywords: physical activity] OR
[Keywords: diet] OR
[Keywords: dietary] OR
[Keywords: sexual health] OR
[Keywords: smoking] OR
[Keywords: sleeping] OR
[Keywords: sedentary] OR
[Keywords: sitting] OR
[Keywords: alcohol] AND
[E-Publication Date: (01/01/2019 to 06/30/2026)]
\end{tcolorbox}

\begin{figure}[b]
    \centering
    \includegraphics[width=0.85\linewidth]{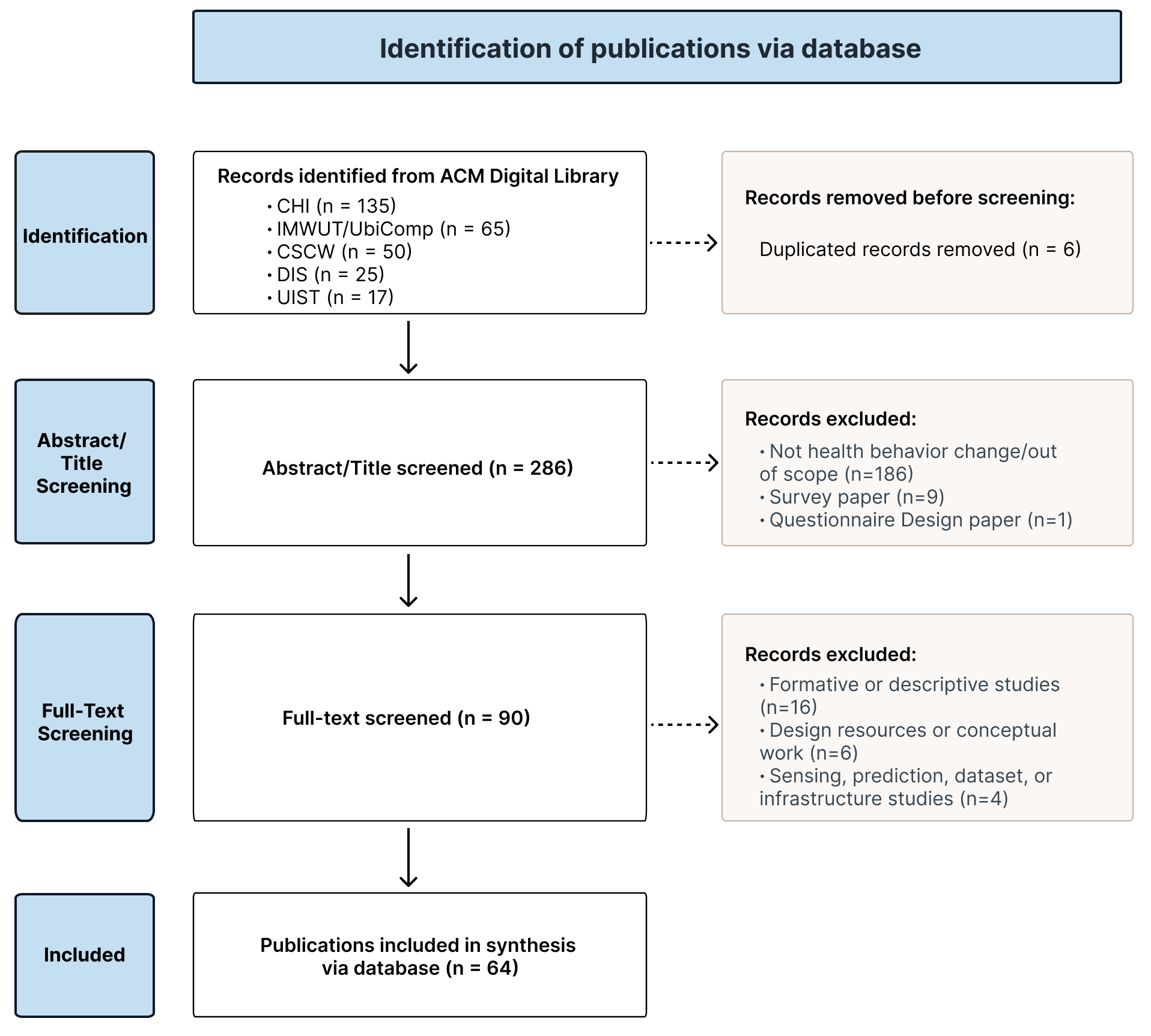}
    \caption{Literature search and screening process, adapted from PRISMA. The ACM Digital Library search identified 292 records, of which 64 publications were retained for BCT coding after deduplication, title and abstract screening, and full-text assessment.}
    \label{fig:candidate_identification}
\end{figure}

\subsection{Paper Screening}

The search yielded 292 records: 135 from CHI, 25 from DIS, 17 from UIST, 65 from IMWUT, and 50 from CSCW. We removed duplicates and screen titles and abstracts to exclude papers outside health behavior change, reviews, and studies focused on measurement scales. These steps leave 90 papers for full-text assessment.
At the full-text stage, we retained papers that implement and evaluate a participant-facing health behavior change intervention. We exclude 26 papers that do not meet this criterion, leaving 64 papers for BCT coding. The literature screening flow in Figure~\ref{fig:candidate_identification} is adapted from PRISMA~\cite{page2021prisma}.

\subsection{BCT Coding and Frequencies}

The first and second authors independently coded the intervention content of all 64 included papers using BCTTv1~\cite{michie2013behavior}. For papers reporting multiple interventions or experimental conditions, coding focuses on the prototype proposed in the paper. After completing the coding, the two authors compare their codes and discuss discrepancies, rechecking the relevant intervention descriptions. The third author, an experienced researcher, makes the final decision on any unresolved cases.
The coding identifies 47 distinct BCTs. Table~\ref{tab:literature_bct_frequencies} reports their paper-level frequencies by BCTTv1 group. Because a paper can contain multiple BCTs, these frequencies are not mutually exclusive.

\begin{table*}[b]
    \centering
    \small
    \setlength{\tabcolsep}{5pt}
    \begin{tabularx}{\textwidth}{
        @{}
        >{\raggedright\arraybackslash}p{0.23\textwidth}
        >{\raggedright\arraybackslash}X
        @{}
    }
        \toprule
        BCTTv1 group & BCTs and paper counts \\
        \midrule
        \midrule
        1. Goals and planning &
        1.1 Goal setting (behavior) ($n=24$);
        1.2 Problem solving ($n=7$);
        1.3 Goal setting (outcome) ($n=3$);
        1.4 Action planning ($n=15$);
        1.5 Review behavior goal(s) ($n=3$);
        1.6 Discrepancy between current behavior and goal ($n=13$);
        1.8 Behavioral contract ($n=1$);
        1.9 Commitment ($n=2$). \\
        \addlinespace[2pt]

        2. Feedback and monitoring &
        2.2 Feedback on behavior ($n=28$);
        2.3 Self-monitoring of behavior ($n=27$);
        2.4 Self-monitoring of outcome(s) of behavior ($n=7$);
        2.7 Feedback on outcome(s) of behavior ($n=6$). \\
        \addlinespace[2pt]

        3. Social support &
        3.1 Social support (unspecified) ($n=14$);
        3.2 Social support (practical) ($n=2$);
        3.3 Social support (emotional) ($n=10$). \\
        \addlinespace[2pt]

        4. Shaping knowledge &
        4.1 Instruction on how to perform a behavior ($n=20$). \\
        \addlinespace[2pt]

        5. Natural consequences &
        5.1 Information about health consequences ($n=7$);
        5.2 Salience of consequences ($n=1$);
        5.4 Monitoring of emotional consequences ($n=5$). \\
        \addlinespace[2pt]

        6. Comparison of behaviour &
        6.1 Demonstration of the behavior ($n=6$);
        6.2 Social comparison ($n=6$). \\
        \addlinespace[2pt]

        7. Associations &
        7.1 Prompts/cues ($n=30$). \\
        \addlinespace[2pt]

        8. Repetition and substitution &
        8.1 Behavioral practice/rehearsal ($n=12$);
        8.2 Behavior substitution ($n=5$);
        8.7 Graded tasks ($n=3$). \\
        \addlinespace[2pt]

        9. Comparison of outcomes &
        9.1 Credible source ($n=1$). \\
        \addlinespace[2pt]

        10. Reward and threat &
        10.1 Material incentive (behavior) ($n=1$);
        10.2 Material reward (behavior) ($n=2$);
        10.3 Non-specific reward ($n=4$);
        10.4 Social reward ($n=6$);
        10.6 Non-specific incentive ($n=1$);
        10.11 Future punishment ($n=1$). \\
        \addlinespace[2pt]

        11. Regulation &
        11.1 Pharmacological support ($n=1$);
        11.2 Reduce negative emotions ($n=7$). \\
        \addlinespace[2pt]

        12. Antecedents &
        12.1 Restructuring the physical environment ($n=8$);
        12.2 Restructuring the social environment ($n=1$);
        12.3 Avoidance/reducing exposure to cues for the behavior ($n=2$);
        12.4 Distraction ($n=1$);
        12.5 Adding objects to the environment ($n=12$). \\
        \addlinespace[2pt]

        13. Identity &
        13.2 Framing/reframing ($n=3$);
        13.5 Identity associated with changed behavior ($n=1$). \\
        \addlinespace[2pt]

        14. Scheduled consequences &
        14.1 Behavior cost ($n=2$);
        14.2 Punishment ($n=2$);
        14.5 Rewarding completion ($n=1$). \\
        \addlinespace[2pt]

        15. Self-belief &
        15.1 Verbal persuasion about capability ($n=6$);
        15.3 Focus on past success ($n=3$);
        15.4 Self-talk ($n=1$). \\

        \bottomrule
    \end{tabularx}
    \caption{The 47 BCTs identified in the 64 included papers, organized by BCTTv1 group~\cite{michie2013behavior}. For each BCT, $n$ denotes the number of papers in which it is identified.}
    \label{tab:literature_bct_frequencies}
\end{table*}

\subsection{Integration of Candidate Sources and Preliminary Selection}
\label{app:bct_selection}

\begin{table*}[t]
    \centering
    \footnotesize
    \setlength{\tabcolsep}{4pt}
    \begin{tabularx}{\textwidth}{
        @{}
        p{0.09\textwidth}
        p{0.24\textwidth}
        X
        @{}
    }
        \toprule
        Code & Source & BCTs \\
        \midrule
        \midrule
        EH &
        HCI research through 2018 summarized by \citet{wang2019health} ($n=10$) &
        1.1 Goal setting (behavior);
        1.4 Action planning;
        1.6 Discrepancy between current behavior and goal;
        2.2 Feedback on behavior;
        2.3 Self-monitoring of behavior;
        3.1 Social support (unspecified);
        4.1 Instruction on how to perform a behavior;
        5.1 Information about health consequences;
        6.2 Social comparison;
        7.1 Prompts/cues. \\
        \addlinespace[2pt]
        RH &
        HCI research since 2019 reviewed in this study ($n=10$) &
        1.1 Goal setting (behavior);
        1.4 Action planning;
        1.6 Discrepancy between current behavior and goal;
        2.2 Feedback on behavior;
        2.3 Self-monitoring of behavior;
        3.1 Social support (unspecified);
        4.1 Instruction on how to perform a behavior;
        7.1 Prompts/cues;
        8.1 Behavioral practice/rehearsal;
        12.5 Adding objects to the environment. \\
        \addlinespace[2pt]
        BS &
        Public-health and behavioral-science corpus ($n=16$) &
        1.1 Goal setting (behavior);
        1.2 Problem solving;
        1.4 Action planning;
        1.5 Review behavior goal(s);
        2.2 Feedback on behavior;
        2.3 Self-monitoring of behavior;
        3.1 Social support (unspecified);
        4.1 Instruction on how to perform a behavior;
        5.1 Information about health consequences;
        6.1 Demonstration of the behavior;
        6.2 Social comparison;
        7.1 Prompts/cues;
        8.1 Behavioral practice/rehearsal;
        8.7 Graded tasks;
        9.1 Credible source;
        12.5 Adding objects to the environment. \\
        \addlinespace[2pt]
        TD &
        Additional candidates proposed through team discussion
        ($n=2$) &
        10.4 Social reward;
        15.3 Focus on past success. \\

        \bottomrule
    \end{tabularx}
    \caption{BCTs contributed by each candidate-identification source.}
    \label{tab:candidate_source_pools}
\end{table*}

Table~\ref{tab:candidate_source_pools} documents the four inputs to candidate identification. For the behavioral-science source, the cutoff accommodated frequency ties and yielded 16 BCTs; including the next frequency group would have substantially enlarged the candidate pool.
The three literature-based pools yielded 17 unique BCTs after deduplication, and team discussion added two further techniques. The ten retained candidates are presented in Table~\ref{tab:candidate_bcts}. Table~\ref{tab:excluded_bct_candidates} documents the remaining nine and their reasons for non-selection. BCTTv1 definitions and coding notes are used to clarify boundaries between related techniques and the requirements of their proposed implementations~\cite{michie2013behavior}.

\begin{table*}[t]
    \centering
    \footnotesize
    \setlength{\tabcolsep}{4pt}
    
    \begin{tabularx}{\textwidth}{
        @{}
        >{\raggedright\arraybackslash}p{0.27\textwidth}
        p{0.10\textwidth}
        >{\raggedright\arraybackslash}X
        @{}
    }
        \toprule
        BCT & Sources & Primary reason for non-selection \\
        \midrule
        \midrule

        2.3 Self-monitoring of behavior &
        EH, RH, BS &
        Photo capture and access to personal photo histories were shared
        across conditions and could already support self-monitoring.
        We therefore treated this BCT as part of the common photo-journaling infrastructure instead of a separate support. \\
        \addlinespace[2pt]
        1.1 Goal setting (behavior) &
        EH, RH, BS &
        Participants were recruited with an existing intention to record,
        so generic goal confirmation offered limited added value.
        Specifying the context, frequency, or duration of capture would
        also involve 1.4 Action planning. \\
        \addlinespace[2pt]
        4.1 Instruction on how to perform a behavior &
        EH, RH, BS &
        All participants required the same instructions about eligible
        meals, capture timing, and valid photographs.
        We therefore treated instruction as common onboarding instead of a separate support. \\
        \addlinespace[2pt]
        8.1 Behavioral practice/rehearsal &
        RH, BS &
        Practice outside actual meals had limited ecological relevance,
        whereas repetition during meals overlapped with repeated
        performance of the study task.
        When intended to associate capture with the meal context, it would
        also involve 8.3 Habit formation. \\
        \addlinespace[2pt]
        12.5 Adding objects to the environment &
        RH, BS &
        BCTTv1 requires objects to be added to the environment and specifies
        that information alone is insufficient.
        The implementations considered by the team required additional
        physical objects and therefore fell outside the smartphone-only scope. \\
        \addlinespace[2pt]
        5.1 Information about health consequences &
        EH, BS &
        Taking or missing a pre-meal photo did not have sufficiently direct
        health consequences to support a credible implementation.
        Linking capture to downstream dietary outcomes would also shift
        attention away from the recording behavior itself. \\
        \addlinespace[2pt]
        6.1 Demonstration of the behavior &
        BS &
        For this simple action, the team could not identify a
        demonstration-only implementation with clear added value.
        Implementations involving practice or instructions would also
        involve BCTs 8.1 and 4.1, respectively. \\
        \addlinespace[2pt]
        9.1 Credible source &
        BS &
        Authoritative endorsement alone offered limited added value for
        pre-meal photo capture.
        Meaningful persuasive content would likely depend on information
        represented by other BCTs, making credible source difficult to
        isolate as a distinct support. \\
        \addlinespace[2pt]
        1.5 Review behavior goal(s) &
        BS &
        This BCT presupposes an existing behavioral goal and reviews it in
        light of achievement.
        A concrete implementation would likely combine goal review with
        discrepancy information or behavioral feedback, limiting its
        distinct role in the candidate set. \\
        \bottomrule
    \end{tabularx}
    \caption{BCTs not retained for the workshop candidate set. EH: HCI research through 2018. RH: HCI research since 2019 reviewed in this study. BS: public-health and behavioral-science corpus.
    Source labels are not mutually exclusive. The decisions reflect the scope and comparative design of this study.}
    \label{tab:excluded_bct_candidates}
\end{table*}

\section{Workshop Procedure and Evaluation Instruments}
\label{app:workshop_evaluation}

\subsection{Detailed Workshop Procedure}
\label{app:workshop_procedure}

\paragraph{Introduction, briefing, and reflection (15 min).}
Participants received an overview of the study and target behavior, then reflected on factors that might support or hinder pre-meal photo capture across meal contexts.

\paragraph{BCT initial screening (20 min).}
Participants reviewed the candidate BCTs and recorded their initial judgments. A BCT was excluded only if all judgments were negative and no participant argued for retaining it. The remaining BCTs proceeded to the design stage.

\paragraph{Individual design (30 min).}
After an overview of basic smartphone functions, each participant developed two distinct smartphone-based support concepts, each based on a different BCT, and considered how the concepts could operate across meal contexts.

\paragraph{Presentation and peer rating (55 min).}
Each participant presented each concept for three minutes, followed by two minutes of Q\&A. Participants then rated each design using a seven-point form covering perceived value, acceptability, burden, smartphone feasibility, and anticipated longer-term use. The complete items are provided in Appendix~\ref{app:peer_rating}.

\paragraph{Group discussion and BCT-level voting (15 min).}
Participants discussed the generated concepts, then judged at the BCT level whether each technique offered a smartphone-based implementation worth further testing to support the formation of a pre-meal photo capture habit.

\paragraph{Follow-up Reflection (15 min).}
After both workshops, we synthesized one standardized support version for each retained BCT from the concepts and discussions. In a reflection phase informed by \citet{ekman2010using}, participants clarified or added points not fully expressed during the workshops, proposed refinements, identified questions for longitudinal evaluation, and nominated three final candidates for the field study. The reflection structure, shared items, and response options are provided in Appendix~\ref{app:followup_reflection}.

\subsection{Peer-Rating Form}
\label{app:peer_rating}
After each presentation, all five participants in the corresponding workshop, including the design author, evaluated the specific support concept using the following 11 items. All items used seven-point response scales. The form also recorded the evaluator, presenter, and BCT represented by the design.

\noindent Q1. I could accept this BCT-based support in real-world
pre-meal settings.

\noindent Q2. This support directly helps users form a habit of
pre-meal photo capture, rather than primarily serving nutrition
analysis or other downstream goals.

\noindent Q3. This support could help pre-meal photo capture even when
users have different ultimate purposes for food-photo journaling.

\noindent Q4. This support could be implemented reliably using a
single smartphone, without additional devices or highly unreliable
context inference.

\noindent Q5. This support would not substantially increase
operational, attentional, or social burden before eating.

\noindent Q6. This support would help users develop their own recording
routine rather than feel controlled or compelled by the system.

\noindent Q7. After approximately two weeks of repeated use, this
support would remain acceptable and helpful.

\noindent Q8. This support would help users establish a routine that
could continue after the support was withdrawn.

\noindent Q9. This design fits routine solo or everyday meals.

\noindent Q10. This design fits time-pressured meals.

\noindent Q11. This design fits social meals.

\subsection{Follow-up Reflection Questionnaire}
\label{app:followup_reflection}

Participants independently evaluated six standardized candidate support versions: Social support, Prompts/cues, Focus on past success, Problem solving, Feedback on behavior, and Social reward. For each version, the questionnaire presented a summary of the workshop ideas, the standardized support synthesized by the research team, and the concerns raised during the workshops. Participants then completed Q1--Q7 for each BCT support before proceeding to the next candidate. Q1--Q5 used seven-point scales, where 1 indicated strong disagreement and 7 indicated strong agreement. Q6 and Q7 were single-choice questions. The questionnaire did not display prior votes, rankings, or scores.

\noindent Q1. This version provides a reasonable way to help people seeking to establish a photo journal complete food-photo capture more often before the first bite.

\noindent Q2. This version could apply to users with different purposes for food-photo journaling, such as reviewing everyday life, understanding eating practices, or preserving personal records.

\noindent Q3. Users could accept using this version continuously for two weeks in everyday life.

\noindent Q4. This version could be delivered reliably and repeatedly using only a smartphone.

\noindent Q5. Examining whether this version remains stable, weakens, or changes during two weeks of support, and whether the behavior is maintained during one subsequent week without support, could yield informative findings.

\noindent Q6. Based on the preceding judgments, how should this candidate proceed? Select one option:
\begin{itemize}
    \item Enter longitudinal testing in its current form.
    \item Enter longitudinal testing after redesign, please state the reason.
    \item Do not enter longitudinal testing, please state the reason.
    \item Insufficient information to judge, please state the reason.
\end{itemize}

\noindent Q7. Which single question about this candidate is most worth examining longitudinally? Select one option:
\begin{itemize}
    \item Whether it helps initiate capture for the current meal.
    \item Whether it is ignored or loses effectiveness over time.
    \item Whether burden increases over time.
    \item Whether it promotes more self-initiated capture.
    \item Whether the behavior persists after support withdrawal.
    \item Whether it applies only to particular goals or contexts.
    \item Whether it depends on another form of support.
    \item Other, please state the question.
\end{itemize}

After evaluating all six candidates, participants completed the following overall reflection.

\noindent FQ1. Select exactly three candidates that would form an informative longitudinal comparison. 

\noindent FQ2. For each selected candidate, anticipate how its role might change during the two-week support period.

\noindent FQ3. For each selected candidate, anticipate what might happen during the subsequent week without support. Select one option:
\begin{itemize}
    \item Mostly maintained.
    \item Partly maintained. 
    \item Clear decline.
    \item Dependent on context.
    \item Uncertain.
\end{itemize}

\noindent FQ4. What difference among the three selected candidates would be most informative to compare? You can consider intervention timing, user burden, dependence on external support, possible loss of effectiveness over time, and behavior after support withdrawal.

\noindent FQ5. You can identify the candidate most in need of modification and suggest one specific adjustment, if you need.

\section{Supporting Quantitative Workshop Results}
\label{app:workshop_results}

This appendix reports the context weights, peer-rating rankings,
BCT-level vote counts, and follow-up rating distributions supporting
Section~\ref{sec:formative_selection}. Values are displayed to at most
two decimal places; calculations used unrounded values.

\subsection{Context Weights}

Each participant's context-importance ratings were normalized to sum to one and used to weight their context-fit ratings, as shown in Table~\ref{tab:context_weights}.

\begin{table*}[h]
\centering
\begin{minipage}[t]{0.43\textwidth}
\vspace{0pt}
\centering
\small
\setlength{\tabcolsep}{3pt}
\begin{tabularx}{\linewidth}{@{}c*{3}{>{\centering\arraybackslash}X}@{}}
\toprule
Participant & Routine & \shortstack{Time-pressured} & Social \\
\midrule
\midrule
P1 & 0.37 & 0.37 & 0.26 \\
P2 & 0.50 & 0.17 & 0.33 \\
P3 & 0.43 & 0.29 & 0.29 \\
P4 & 0.33 & 0.33 & 0.33 \\
P5 & 0.40 & 0.33 & 0.27 \\
\addlinespace
P6 & 0.38 & 0.15 & 0.46 \\
P7 & 0.37 & 0.37 & 0.26 \\
P8 & 0.33 & 0.40 & 0.27 \\
P9 & 0.39 & 0.39 & 0.22 \\
P10 & 0.35 & 0.35 & 0.30 \\
\bottomrule
\end{tabularx}
\caption{Normalized context-importance weights.}
\label{tab:context_weights}
\end{minipage}\hfill
\begin{minipage}[t]{0.54\textwidth}
\vspace{0pt}
\centering
\small
\setlength{\tabcolsep}{3pt}
\begin{tabularx}{\linewidth}{@{}
>{\raggedright\arraybackslash}X
>{\centering\arraybackslash}p{0.13\linewidth}
>{\centering\arraybackslash}p{0.13\linewidth}@{}}
\toprule
BCT & \shortstack{W1\\Yes/No} & \shortstack{W2\\Yes/No} \\
\midrule
\midrule
Social support & 3/0 & 4/1 \\
Focus on past success & 2/0 & 4/0 \\
Prompts/cues & 4/0 & 1/3 \\
Feedback on behavior & 0/3 & 3/1 \\
Problem solving & 0/1 & 3/0 \\
Social reward & 4/0 & \textemdash{} \\
Action planning & 1/1 & 0/2 \\
Graded tasks & 0/0 & \textemdash{} \\
Discrepancy between current behavior and goal & \textemdash{} & \textemdash{} \\
Social comparison & \textemdash{} & \textemdash{} \\
\bottomrule
\end{tabularx}
\caption{BCT-level Yes/No judgments in W1 and W2.}
\label{tab:complete_bct_judgments}
\end{minipage}
\end{table*}

\subsection{BCT-Level Judgments}

Table~\ref{tab:complete_bct_judgments} reports the final Yes/No counts for all ten candidates. Blank entries at a reached voting stage were coded as zero. An em-dash indicates that the BCT did not reach that stage.

\subsection{Peer-Rating Results}

We ranked the 20 concepts separately with and without their own self-ratings. For each of the nine dimensions, we calculated each concept's median rating and percentile rank, averaging tied ranks. Concepts with more dimensions above the 50th percentile ranked higher. If tied, we compared counts above the 75th percentile and, if still tied, the median of the nine dimension medians, favoring higher values. The top half contained ten concepts in each ranking, with no ties across the cutoff. Table~\ref{tab:peer_rating_rankings} lists all concepts by BCT and designer, ordered by the ranking including self-ratings.

\begin{table*}[t]
\centering
\small
\setlength{\tabcolsep}{5pt}
\begin{tabularx}{\textwidth}{@{}
>{\raggedright\arraybackslash}X
>{\centering\arraybackslash}p{0.15\textwidth}
>{\centering\arraybackslash}p{0.3\textwidth}
>{\centering\arraybackslash}p{0.3\textwidth}@{}}
\toprule
BCT & Designer & \shortstack{Pooled rank including self-ratings}
& \shortstack{Pooled rank excluding self-ratings} \\
\midrule
\midrule
Focus on past success & P2 & \textbf{1} & \textbf{2} \\
Focus on past success & P9 & \textbf{2} & \textbf{6} \\
Graded tasks & P4 & \textbf{3} & \textbf{4} \\
Social support & P4 & \textbf{4} & \textbf{1} \\
Social reward & P3 & \textbf{5} & \textbf{2} \\
Action planning & P2 & \textbf{5} & \textbf{5} \\
Social reward & P1 & \textbf{7} & \textbf{9} \\
Feedback on behavior & P5 & \textbf{8} & \textbf{7} \\
Focus on past success & P10 & \textbf{8} & 12 \\
Prompts/cues & P1 & \textbf{10} & \textbf{8} \\
Feedback on behavior & P9 & 11 & 11 \\
Social support & P6 & 12 & 15 \\
Problem solving & P3 & 13 & 14 \\
Social support & P7 & 14 & 13 \\
Problem solving & P7 & 15 & \textbf{10} \\
Feedback on behavior & P10 & 16 & 19 \\
Problem solving & P8 & 16 & 16 \\
Problem solving & P5 & 18 & 18 \\
Action planning & P8 & 19 & 20 \\
Prompts/cues & P6 & 20 & 17 \\
\bottomrule
\end{tabularx}
\caption{Peer-rating ranks for all 20 concepts pooled across both
workshops. Bold ranks indicate the top half. Tied concepts share the
same rank.}
\label{tab:peer_rating_rankings}
\end{table*}

\subsection{Follow-up Ratings}

\begin{figure*}[h]
\centering
\includegraphics[width=.9\textwidth]{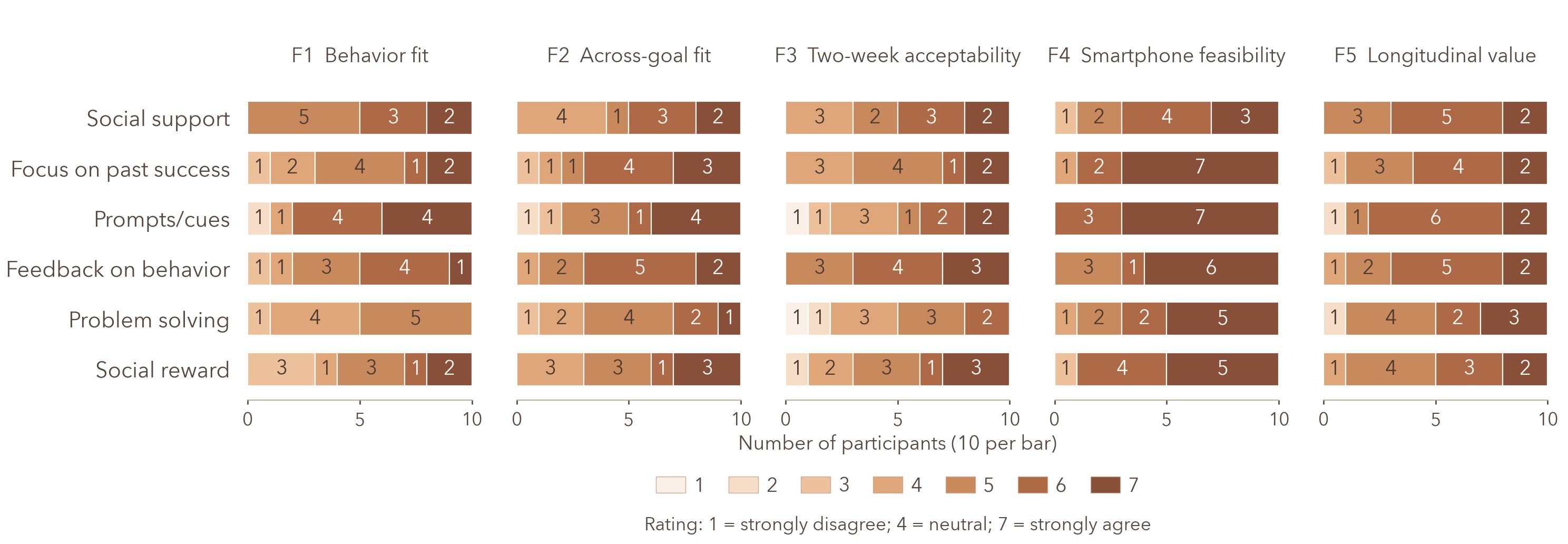}
\caption{Workshop follow-up ratings of the six standardized support candidates ($N=10$). Each bar contains ten responses, and segment labels indicate participant counts. F1: fit with pre-meal photo capture; F2: applicability across recording goals; F3: acceptability over two weeks; F4: smartphone feasibility; F5: information value of longitudinal evaluation. Ratings range from 1 (strongly disagree) to 7 (strongly agree).}
\label{fig:followup_rating_distributions}
\end{figure*}

Figure~\ref{fig:followup_rating_distributions} shows all responses to the five seven-point agreement items for each standardized support version from 10 participants.

\section{Supplementary Results from the Longitudinal Study}
\label{app:longitudinal_study}

\begin{figure}[h]
    \centering
    \includegraphics[width=0.85\linewidth]{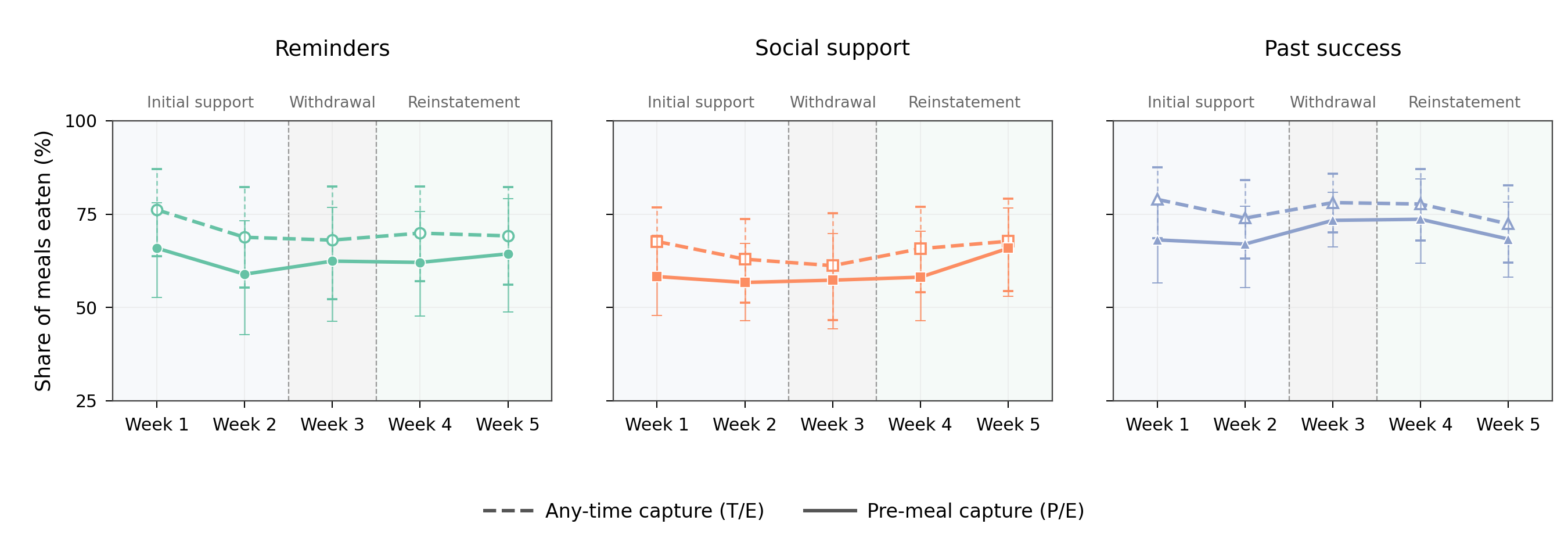}
    \caption{Weekly capture coverage regardless of timing ($T/E$, dashed lines) and confirmed pre-meal capture adherence ($P/E$, solid lines) by support condition. Values are group means of participant-level proportions, with meals eaten as the denominator for both measures.
    }
    \label{fig:weekly_capture_rates_by_support}
\end{figure}

This appendix provides supplementary views of the longitudinal results reported in Sections~\ref{sec:trajetory_and_changes}. 

\begin{figure}[h]
    \centering
    \includegraphics[width=.85\linewidth]{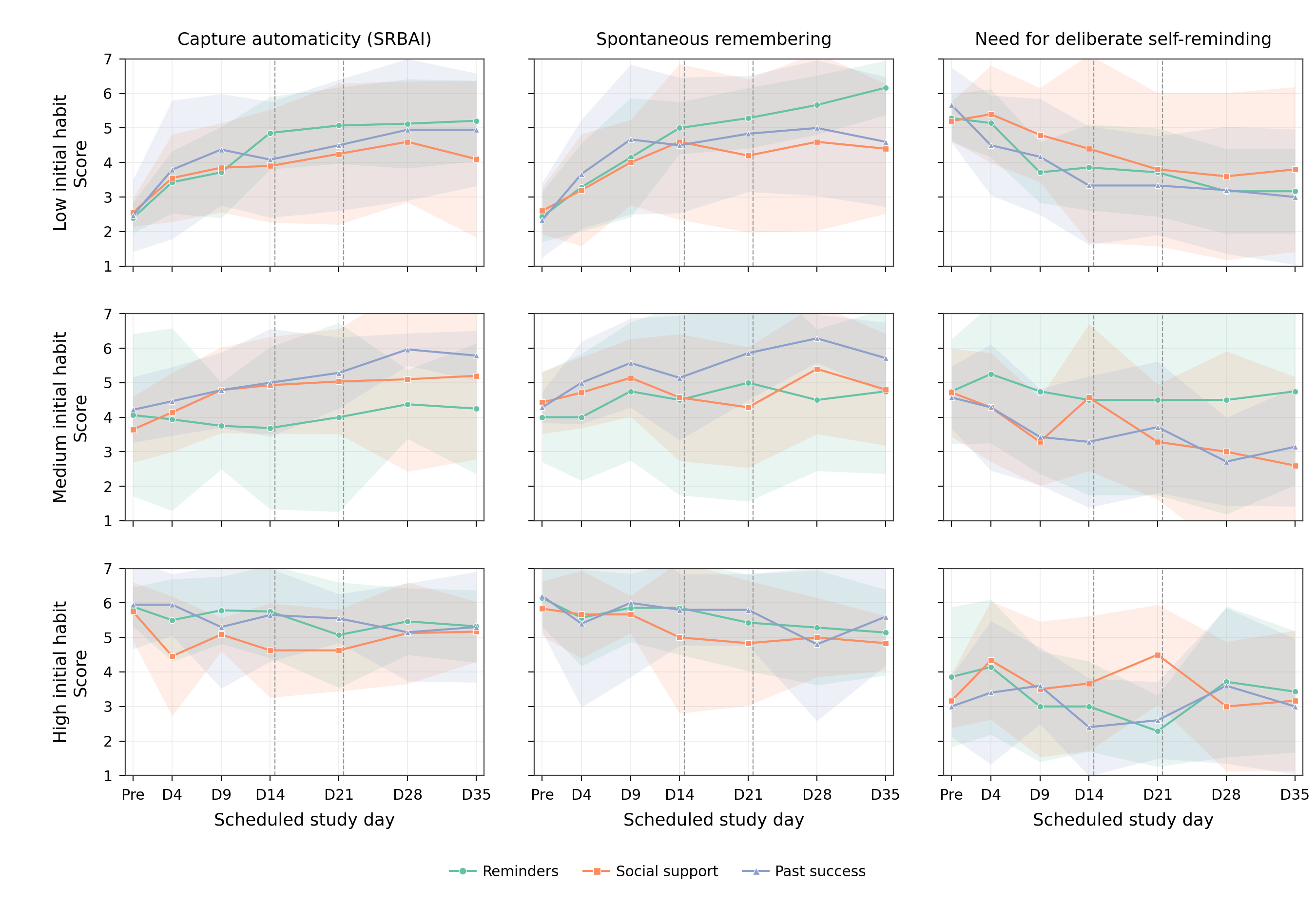}
    \caption{Subjective habit trajectories by support condition within each fixed initial-habit stratum. Rows show low, medium, and high baseline habit levels. Columns show automaticity, spontaneous remembering, and the need for deliberate self-reminding.}
    \label{fig:composite_questionnaire_strata}
\end{figure}

Figure~\ref{fig:weekly_capture_rates_by_support} compares overall capture coverage with pre-meal capture adherence. 
Figure~\ref{fig:composite_questionnaire_strata} presents subjective habit trajectories by support condition and fixed initial-habit stratum. Stratum membership is retained from baseline, and the paired follow-up comparisons include participants with observations at both time points.

\section{Gen AI Use}
We used generative AI to create selected visual elements in Figure~\ref{fig:teaser} and Figure~\ref{fig:study_procedure}. We employed it to polish the manuscript's language and enhance its overall fluency.